\documentclass[twocolumn,english,prl, superscriptaddress]{revtex4-1}
\usepackage[T1]{fontenc}
\usepackage[latin9]{inputenc}
\usepackage{color}
\usepackage{babel}
\usepackage{amsmath}
\usepackage{amssymb}
\usepackage{graphicx}
\usepackage{wasysym}
\usepackage[unicode=true,pdfusetitle,
 bookmarks=true,bookmarksnumbered=false,bookmarksopen=false,
 breaklinks=false,pdfborder={0 0 1},backref=false,colorlinks=true]
 {hyperref}

\makeatletter
\usepackage{braket}
\usepackage{wrapfig}
\renewcommand{\p@subsection}{\thesection.}
\hypersetup{colorlinks,citecolor=blue}

\makeatother

\begin{document}
\title{Renormalized Lindblad Driving: A Numerically-Exact Nonequilibrium
Quantum Impurity Solver}
\author{Matan Lotem}
\affiliation{Raymond and Beverly Sackler School of Physics and Astronomy, Tel Aviv
University, Tel Aviv 6997801, Israel}
\author{Andreas Weichselbaum}
\affiliation{Department of Condensed Matter Physics and Materials Science, Brookhaven
National Laboratory, Upton, NY 11973-5000, USA}
\affiliation{Arnold Sommerfeld Center for Theoretical Physics, Center for NanoScience,
and Munich Center for Quantum Science and Technology, Ludwig-Maximilians-Universität
München, 80333 Munich, Germany}
\author{Jan von Delft}
\affiliation{Arnold Sommerfeld Center for Theoretical Physics, Center for NanoScience,
and Munich Center for Quantum Science and Technology, Ludwig-Maximilians-Universität
München, 80333 Munich, Germany}
\author{Moshe Goldstein}
\affiliation{Raymond and Beverly Sackler School of Physics and Astronomy, Tel Aviv
University, Tel Aviv 6997801, Israel}
\begin{abstract}
The accurate characterization of nonequilibrium strongly-correlated
quantum systems has been a longstanding challenge in many-body physics.
Notable among them are quantum impurity models, which appear in various
nanoelectronic and quantum computing applications. Despite their seeming
simplicity, they feature correlated phenomena, including small emergent
energy scales and non-Fermi-liquid physics, requiring renormalization
group treatment. This has typically been at odds with the description
of their nonequilibrium steady-state under finite bias, which exposes
their nature as open quantum systems. We present a novel numerically-exact
method for obtaining the nonequilibrium state of a general quantum
impurity coupled to metallic leads at arbitrary voltage or temperature
bias, which we call ``RL-NESS'' (Renormalized Lindblad-driven NonEquilibrium
Steady-State). It is based on coherently coupling the impurity to
discretized leads which are treated exactly. These leads are furthermore
weakly coupled to reservoirs described by Lindblad dynamics which
impose voltage or temperature bias. Going beyond previous attempts,
we exploit a hybrid discretization scheme for the leads together with
Wilson's numerical renormalization group, in order to probe exponentially
small energy scales. The steady-state is then found by evolving a
matrix-product density operator via real-time Lindblad dynamics, employing
a dissipative generalization of the time-dependent density matrix
renormalization group. In the long-time limit, this procedure successfully
converges to the steady-state at finite bond dimension due to the
introduced dissipation, which bounds the growth of entanglement. We
thoroughly test the method against the exact solution of the noninteracting
resonant level model. We then demonstrate its power using an interacting
two-level model, for which it correctly reproduces the known limits,
and gives the full $I$-$V$ curve between them.
\end{abstract}
\maketitle
\global\long\def\k#1{\Ket{#1}}%
\global\long\def\b#1{\Bra{#1}}%
\global\long\def\bk#1{\Braket{#1}}%

\section{Introduction}

Quantum impurity models have fascinated theoreticians for several
decades. These models seem extremely simple -- they describe a small,
typically interacting, quantum system, i.e., the impurity, coupled
to a non-interacting environment. The quantum impurity consists of
only a few degrees of freedom, so that its spectrum can be obtained
exactly. However once this interacting impurity is coupled to the
seemingly innocent quadratic environment, it gives rise to highly
correlated behavior and exotic phenomena which cannot be explained
solely in terms of the bare impurity, such as the Kondo effect (including
its non-Fermi-liquid multichannel varieties) \citep{Kondo64,hewson_1993}.
Quantum impurities can thus be seen as the basic building blocks of
higher-dimensional strongly-interacting systems. The most striking
feature of these arising phenomena is that they can occur at emergent
energy scales which, a priori, do not appear in the Hamiltonian of
either the bare impurity or the environment. An example is the Kondo
temperature, which can be smaller by several orders of magnitude than
any bare energy scale. Thus, in order to expose the physics of these
models, they must be analyzed in a renormalization group (RG) framework.
As of today, the most successful method for treating such problems
in or close to equilibrium, is Wilson's numerical renormalization
group (NRG) \citep{Wilson75,Bulla08}, a numerically exact RG procedure
for integrating out high-energy modes and probing arbitrarily small
energy scales.

A wide range of devices with various nanoelectronic and quantum computing
applications, including semiconductor quantum dots \citep{Goldhaber-Gordon1998,Cronenwett540},
carbon nanotubes coupled to metallic leads \citep{Nygard2000,Buitelaar02},
and molecular junctions \citep{Park2002,Liang2002}, can be described
as quantum impurity models, with the environment corresponding to
two macroscopic leads. Most of the their applications involve imposing
a voltage (chemical potential) or temperature bias between the leads
will result in a nonequilibrium steady-state (NESS), with a tunneling
current flowing through the impurity. Experimental results for such
systems have successfully been explained in different limits, e.g.,
by linear response theory together with equilibrium NRG for small
bias, or by solving a master equation at large temperature or voltage
bias \citep{Beenakker91}. However, for arbitrary bias, a quantitative
theoretical description of the NESS properties is still an open challenge.
Any complete solution to this problem must (i) capture interaction
induced many-body correlations, (ii) resolve a wide range of energy
scales, and (iii) deal with an open system at its steady-state. 

Attempts to generally tackle this problem analytically, e.g., in an
RG framework \citep{Rosch03,Kehrein05,Pletyukhov12} or by Keldysh
field integral formulation \citep{Smirnov13} are so far restricted
to uncontrolled approximations. Bethe ansatz approaches have also
been tried \citep{Mehta06,culver2019nonperturbative}, but are typically
case-specific. Therefore much focus has been placed on finding a general
numerical solution. A class of such attempts is based on capturing
the many-body correlations by modeling the environment as large (but
finite) leads, and evolving the many-body state of this finite system
towards a finite-time quasi-steady-state, e.g., using the time-dependent
density matrix renormalization group (tDMRG) method \citep{Boulat08,daSilva08,Eckel_2010}.
This approach has further been extended by treating the finite leads
as open systems, governed by Lindblad dynamics, and similarly evolving
in time towards a well defined steady-state \citep{Dorda15,brenes2019tensornetwork}.
The Lindblad approach has also been recently investigated in the context
of density functional theory \citep{Hod16}. However, these attempts
are typically limited in terms of the range of energy scales explored
by the finite number of energy levels in the leads, with no RG procedure
exploited in order to integrate out high-energy modes. Other numerically
exact approaches applied to this problem are reported in \citep{Tanimura89,YAN2004216,Wang_2008,Cohen14},
but are also not designed to explore the wide range of energy scales.
Two attempts to leverage the unrivaled success of NRG in equilibrium
and extend it out of equilibrium, are the so-called scattering-states
NRG \citep{Anders08}, and the NRG-tDMRG scheme \citep{Schwarz18},
with the latter a predecessor of the method presented in this paper.
These attempts have been quite successful at resolving a wide range
of energy scales, while also capturing the many-body correlations.
Yet while the former is plagued by logarithmic discretization artifacts
within the dynamical energy window, the latter is based on non-dissipative
time evolution of a finite, and thus closed system, which results
in a quasi-steady-state in a limited time interval, making it challenging
to extract steady-state observables. 

In this work we present a novel algorithm combining the full power
of NRG and tDMRG for capturing many-body correlations at a wide range
of energy scales, together with open system dynamics, in order to
obtain an actual nonequilibrium steady-state. In what follows, we
will refer to this approach as the Renormalized Lindblad-driven NESS
(RL-NESS) method. The starting point of the presented method is a
general impurity coupled to continuous leads (i.e., leads with a continuous
spectrum). As shown in Fig.~\ref{fig:Mappings}, each continuous
lead is separated into a finite set of representative discrete energy
levels, which in turn are coupled to the remaining continuous modes.
The impurity together with this finite set of energy levels (large
enough to allow the coherent formation of, e.g., the Kondo screening
cloud, and the emergence of energy scales such as the Kondo temperature),
is considered as an open system, coupled to an environment consisting
of the remaining continuous modes, which are traced out. The latter
is performed under the Born and Markov approximations, i.e., that
the environment is memory-less and indifferent to the state of the
system. As a result the dynamics of the system is governed by a Lindblad
\citep{Breuer_2007} master equation:

\begin{equation}
\frac{d\rho}{dt}=\mathcal{L}\rho=-i\left[H,\rho\right]+\mathcal{D}\rho.\label{eq:Lindbald-equation}
\end{equation}
The Liouvillian super-operator $\mathcal{L}$ can be separated into
a von Neumann term consisting of the discrete system Hamiltonian $H$,
and a dissipative super-operator $\mathcal{D}$, describing a suitably-modeled
dissipation into the environment. The two key elements of our method
are: (i) the specific choice of discrete energy levels, such that
high-energy modes can be integrated out, and (ii) the numerical solution
of the Lindblad equation in the low-energy dynamical regime, formulated
as a tensor-network algorithm.

With these requirements in mind, the Lindblad equation is obtained
as follows: The Hamiltonian of the discrete system is derived by employing
a mixed discretization scheme that crosses over from logarithmic to
linear level spacing at the bias scale \citep{Schwarz18}. This permits
integrating out modes whose energies are high compared to the bias
voltage or temperature by means of NRG, with the logarithmic RG flow
eventually cut off at this scale. Instead of formally deriving the
dissipators, they are chosen based on two criteria: (i) the solution
of the Lindblad equation reproduces the continuum limit, and (ii)
Eq.~(\ref{eq:Lindbald-equation}) can be numerically solved efficiently.
An important property of the chosen dissipators is that they are local
in the basis in which the leads are diagonal. A set of exact transformations,
dubbed the Lindblad driven discretized leads (LDDL) scheme \citep{Schwarz16},
is then applied to the Lindblad equation, mapping it to a so-called
chain geometry, which, due the short-rangedness (or locality) of interactions
is more favorable for treatment in the tensor-network framework, e.g.,
by tDMRG \citep{White92,Schollwoeck11}. At this stage, high-energy
modes (far above the bias voltage and temperature scales) are integrated
out using \emph{equilibrium} NRG, arriving at a local Lindblad equation
in an effective low-energy basis. The state of the system is represented
as a matrix-product density operator (MPDO), and is evolved in real
time by a dissipative variant of tDMRG in Liouvillian space until
convergence to a steady-state is obtained. Due to the dissipation
induced by the environment, the entanglement entropy of the system
saturates as function of time, rather than diverging, as is the case
in the absence of dissipation. Hence the long-time limit steady-state
can be obtained with finite MPDO bond dimension. A full description
of the method will be presented in Sec.~\ref{sec:Method}.

By repeating the simulations for different bias voltages, a full $I$-$V$
curve can be obtained. When numerically differentiated, one obtains
the differential conductance. The method is demonstrated on two spinless
fermionic models: the non-interacting resonant level model (RLM),
and an interacting two-level model (I2LM). The RLM, discussed in Sec.~\ref{sec:Error-Analysis}
can be solved exactly in the single particle basis (in and out of
equilibrium). It will therefore serve as a benchmark for the presented
method. The I2LM, discussed in Sec.~\ref{sec:Interacting-System},
contains two interacting dot levels with level spacing $\Delta$ and
interaction energy $U$. Our method recovers known results in the
limits of small and large bias, yet goes beyond them by giving the
full $I$-$V$ curve. Conclusions and future directions are discussed
in Sec.~\ref{sec:Discussion}, followed by a series of appendixes
covering technical details.

\section{Method\label{sec:Method}}

\begin{figure*}[t]
\begin{centering}
\includegraphics[width=1\textwidth]{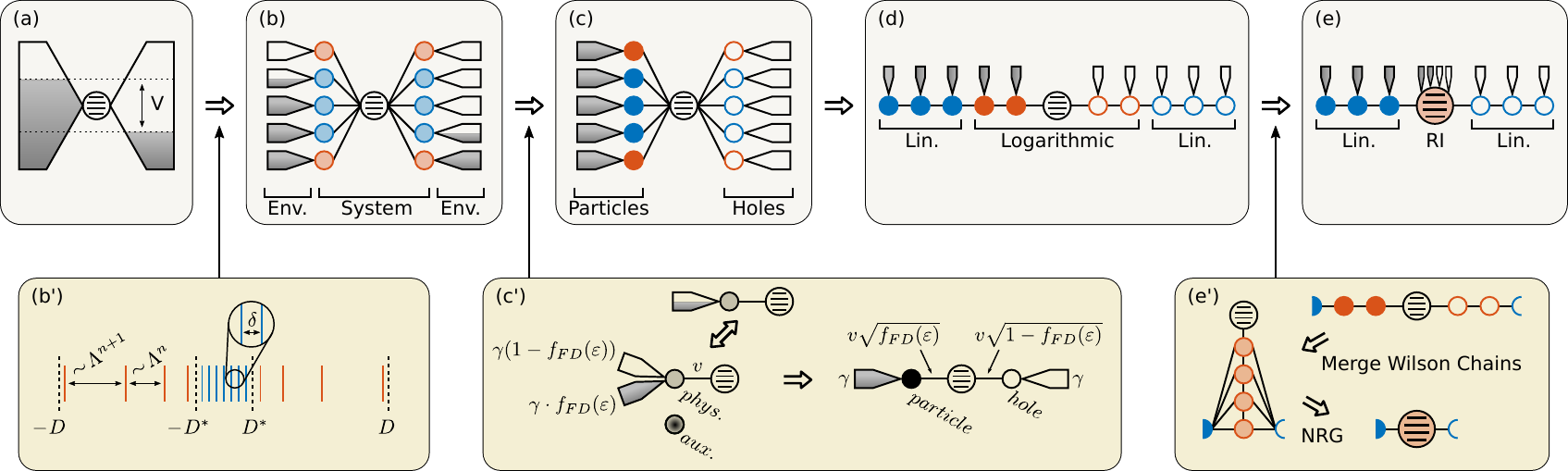}
\par\end{centering}
\caption{Schematic description of the RL-NESS method leading up to the point
of solving the Lindblad equation. (a) The system of interest is a
general impurity coupled to two macroscopic and thus continuous leads
at chemical potential difference $V$. This system can be mapped exactly
onto (b), where the bath has been coarse-grained into distinct intervals.
Each of these is written as a representative level that the impurity
couples to, and a continuous bath consisting of the remainder of states
in that interval. The discrete set of energy levels together with
the impurity then forms a finite system. The remaining lead modes
are integrated out, resulting in a Lindblad bath coupled to each discrete
energy level. (b') The width of the intervals is chosen according
to a logarithmic-linear discretization scheme, such that levels in
the low-energy window $\left[-D^{*},+D^{*}\right]$ are equally spaced
by $\delta$ (blue), with a smooth transition to logarithmic spacing
$\sim\Lambda^{n}$ at large energies (red). (c) The targeted occupation
of each individual lead mode, originally encoded in the couplings
to the Lindblad baths in (b), can be transferred onto the lead-impurity
couplings, such that the resulting two leads now represent particles
or holes, and are driven to be completely full or empty respectively.
(c') This rotation of the local Liouvillian basis is performed separately
for each original (physical) lead level. A general such level is coupled
to the impurity with coupling constant $v$, and to two Lindblad baths,
one filling it and the other emptying it, at rates proportional to
$\gamma$ and chosen such that they drive the level towards its equilibrium
occupation (determined by chemical potential and temperature). From
the Liouvillian description, an auxiliary level is introduced at the
same energy, and linear combinations of the two levels are chosen
such that one is driven to be completely full (particle) and the other
to be completely empty (hole), both at rate $\gamma$. (d) The particle
and hole leads can be exactly mapped onto nearest-neighbor Wilson
chains via tridiagonalization, with local dissipators filling one
chain and emptying the other. The hopping amplitudes along the chains
away from the impurity initially exhibit exponential decay due to
the logarithmic discretization at large energies (red), until they
cross over into more uniform hopping amplitudes of order $\delta$
in the linear discretization regime (blue). (e) The sites in the logarithmic
sector, together with the impurity, are numerically integrated out
in standard NRG spirit. This provides an effective subspace for the
low-energy description in terms of an effective renormalized impurity
(RI) with multiple dissipators. (e') This is achieved by collecting
all the logarithmic sector sites into a single Wilson chain (via a
re-tridiagonalization) for the sake of numerical stability of the
subsequent iterative diagonalization by NRG. The fixed number of states
coming out of the last NRG iteration constitute the RI low-energy
subspace. \label{fig:Mappings}}
\end{figure*}

In this section the RL-NESS method is outlined in detail. The initial
part follows much of the strategy in Ref.~\citep{Schwarz18}. We
start by presenting a general impurity model with continuous leads
in Sec.~\ref{subsec:Model}. The leads are then discretized in Sec.~\ref{subsec:Lindblad-Equation},
resulting in a Lindblad equation for a discrete system. In Sec.~\ref{subsec:Local-Form}
we follow by a short overview of the LDDL scheme, used to bring this
equation into a local form, both in the Hamiltonian and in the dissipators.
In Sec.~\ref{subsec:Renormalized-Impurity} we integrate out high-energy
modes by NRG in order to obtain a renormalized impurity. In Sec.~\ref{subsec:MPDO-Solution}
we describe a matrix-product density operator procedure for real-time
evolution towards the steady-state. Finally, in Sec.~\ref{subsec:Observables}
we discuss the extraction of observables from the obtained steady-state.
Steps B-D are described schematically in Fig.~\ref{fig:Mappings},
and step E is described in Fig.~\ref{fig:MPDO}. Throughout this
section, super-operators acting on the density matrix will be represented
in calligraphic script, while regular operators will be represented
in Roman script. Tensor-network calculations (NRG, MPDO evolution)
were implemented using the QSpace tensor library, which can exploit
both abelian and non-abelian symmetries on a generic footing \citep{Wb12,Wb12tns}.

\subsection{Model\label{subsec:Model}}

The total Hamiltonian of an impurity system can be generically separated
into three parts: the (interacting) impurity, the non-interacting
leads with a continuum density of states, and the coupling between
them:

\begin{equation}
H_{\mathrm{total}}=H_{\mathrm{dot}}+H_{\mathrm{coupling}}+H_{\mathrm{leads}}.\label{eq:general-Hamiltonian}
\end{equation}
The dot Hamiltonian is given by $m$ (here spinless) levels $\lambda$
with onsite Coulomb repulsion $U$:
\begin{equation}
H_{\mathrm{dot}}=\sum_{\lambda=1}^{m}\varepsilon_{\lambda}n_{d\lambda}+\tfrac{U}{2}n_{d}\left(n_{d}-1\right),\label{eq:general-dot-Hamiltonain}
\end{equation}
with fermionic creation operators $d_{\lambda}^{\dagger}$, and total
impurity occupation $n_{d}=\sum_{\lambda}n_{d\lambda}$, where $n_{d\lambda}=d_{\lambda}^{\dagger}d_{\lambda}$.
More complicated local interactions, such as exchange interactions
or spin Hund's coupling, may also be incorporated.

The lead Hamiltonian in this work is described by two metallic, i.e.,
non-interacting leads located left and right of the impurity. They
are assumed to be featureless, with constant hybridization $\Gamma_{\alpha\lambda}$
of lead $\alpha\in\left\{ L,R\right\} $ with the impurity level $\lambda\in\left\{ 1,\ldots,m\right\} $
over a bandwidth $\varepsilon\in\left[-D,+D\right]$, resulting in
the total hybridization strength $v_{\alpha\lambda}\equiv\sqrt{\frac{2D\cdot\Gamma_{\alpha\lambda}}{\pi}}$.
The lead and coupling Hamiltonians can therefore be written in the
diagonal bath basis as:
\begin{align}
H_{\mathrm{leads}} & =\sum_{\alpha}\int_{-D}^{D}d\varepsilon\,\varepsilon\thinspace c_{\alpha\varepsilon}^{\dagger}c_{\alpha\varepsilon},\label{eq:CL-H-leads}\\
H_{\mathrm{coupling}} & =\sum_{\alpha\lambda}v_{\alpha\lambda}\underbrace{\int_{-D}^{D}\tfrac{d\varepsilon}{\sqrt{2D}}\Bigl(c_{\alpha\varepsilon}^{\dagger}}_{\equiv c_{\alpha0}^{\dagger}}d_{\lambda}+\mathrm{H.c.}\Bigr),\label{eq:CL-H-coupling}
\end{align}
where $c_{\alpha\varepsilon}^{\dagger}$ creates an electron in lead
$\alpha$ at energy $\varepsilon$. As indicated, $c_{\alpha0}^{\dagger}$
defines the normalized bath level that the impurity couples to, i.e.,
at the location of the impurity, obeying $\{c_{\alpha0},c_{\alpha0}^{\dagger}\}=1$.
The generalization to spinfull and multi-channel leads is straightforward,
while the generalization to a featured hybridization function is conceptually
also possible. Throughout, we assume the limit of large bandwidth,
i.e., that all energy scales and parameters are much smaller than
$D$. Without loss of generality then, the voltage bias is chosen
symmetric with respect to the Fermi energy, so that the chemical potentials
of the leads are $\mu_{L}=-\mu_{R}=-\frac{V}{2}$ (taking unit of
charge $e=1$, throughout). For concreteness we will mostly concentrate
on the case of zero temperature in both leads, but the described procedure
also applies to finite and non-equal temperatures. 

\subsection{Lindblad Equation\label{subsec:Lindblad-Equation}}

The Lindblad equation is a first-order linear differential equation.
Its general solution, given some initial condition $\rho_{0}$, can
thus be written by exponentiating the Liouvillian super-operator:
\begin{equation}
\frac{\partial\rho}{\partial t}=\mathcal{L}\rho\quad\Rightarrow\quad\rho\left(t\right)=e^{\mathcal{L}t}\rho_{0}.\label{eq:Lindblad-evolution}
\end{equation}
The dynamics in our case is designed to have a unique nonequilibrium
steady-state defined by 
\begin{equation}
\mathcal{L}\rho_{\mathrm{NESS}}\equiv0\quad\iff\quad\rho_{\mathrm{NESS}}=\underset{t\rightarrow\infty}{\lim}\rho\left(t\right),\label{eq:Lindblad-NESS}
\end{equation}
i.e., either as a solution of a linear equation (l.h.s.), or as the
state to which an arbitrary initial states decays to in the long-time
limit (r.h.s).

The first stage in the RL-NESS method is obtaining a Lindblad equation
for a discrete system from the original continuous system, as shown
in Fig.~\ref{fig:Mappings}(b). Formally, this can be done by dividing
the full band $\left[-D,+D\right]$ of each lead into consecutive
distinct intervals $I_{n}$. By the bilinear structure of the coupling
in Eq.~(\ref{eq:CL-H-coupling}), the impurity couples to a particular
state in each interval which itself then is coupled to the remainder
of the states in that interval. The latter can be integrated out,
leaving a single representative level for each interval. Explicitly
performing this integration (under the Born and Markov approximations)
yields the structure of the Lindblad baths and their couplings to
the system. However, we will allow ourselves some freedom in choosing
the exact values of the couplings to the Lindblad baths so as to simplify
the subsequent simulation of the driven system, while enforcing that
the correct steady-state is obtained.

The choice of the intervals $I_{n}$ and corresponding energy levels
relates to coarse-graining that depends on a discretization scheme.
A common discretization scheme used for treating quantum impurity
models is the logarithmic discretization scheme, introduced by Wilson
as part of the NRG \citep{Wilson75,Bulla08}. This scheme produces
discrete semi-infinite leads with level spacing shrinking exponentially
as the lead Fermi energy is approached. It is designed to generate
energy scale separation, and subsequently justifies integrating out
of high-energy modes via iterative exact diagonalization as part of
a logarithmic RG flow. This enables us to accurately and reliably
resolve exponentially small energy scales which frequently arise in
impurity models due to Kondo-like physics. However, for an open system,
e.g., via coupling to a thermal reservoir or the presence of finite
voltage bias, energy scale separation ceases to exist below the corresponding
energy scale, and the logarithmic RG flow will be cut off. In the
nonequilibrium case this gives rise to a dynamical low-energy window
described by a reduced bandwidth $D^{\ast}$, which is of order of
the bias voltage or temperature (see below). For a least-biased numerical
approach then, the discretization scheme within this regime should
be uniform. Therefore, RL-NESS employs a mixed discretization scheme
\citep{Guettge13,Schwarz18}. This consists of a logarithmically discretized
region extending from the band edge down to just above the lead bias
voltage or temperature, that smoothly crosses over into a linearly
discretized region (with uniform level spacing) in the bias window
$\left[-D^{\ast},+D^{\ast}\right]$. Such a scheme allows one to make
use of NRG to integrate out high-energy modes (relative to $V$ or
$T$), in order to obtain an effective low-energy nonequilibrium system,
to be simulated in a controlled manner by a DMRG-like approach. This
scheme has also been discussed for the setup of two leads with a voltage
or temperature bias in a previous work \citep{Schwarz18}, but without
the Lindblad driving (previously suggested in Ref.~\citep{Schwarz16}).
It is therefore briefly outlined here for completeness.

We define $D^{*}$, the characteristic energy scale of the leads,
as the energy at which the Fermi-Dirac distribution of the lead drops
below some pre-selected threshold. For zero temperature this implies
$D^{*}=\frac{V}{2}$, while for finite temperature the specific value
of $D^{*}$ depends on the chosen threshold. The intervals $I_{n}$,
as shown in Fig.~\ref{fig:Mappings}(b'), are chosen such that in
the range $\left[-D^{*},+D^{*}\right]$ they are of equal size $\delta$,
referred to as the linear discretization parameter, while away from
this range they scale exponentially as $\sim\Lambda^{n}$, where $\Lambda>1$
is referred to as the logarithmic discretization parameter. In the
intermediate region the interval widths cross over smoothly between
being constant and growing exponentially. In each interval a representative
energy level $\varepsilon_{n}$ is selected with corresponding coupling
$v_{\alpha n\lambda}$ to the impurity $\lambda$th level. For details
regarding the choice the intervals and corresponding energies and
couplings see Appendix \ref{sec:Appendix-Lin-Log-Disc}. The same
intervals are chosen for both leads such that by construction $\varepsilon_{n}$
are lead independent, while the coupling constants $v_{\alpha n\lambda}$
can differ between the leads. The resulting leads and coupling Hamiltonians
are: 
\begin{align}
H_{\mathrm{leads}}^{\mathrm{(disc)}} & =\sum_{\alpha,n}\varepsilon_{n}c_{\alpha n}^{\dagger}c_{\alpha n},\label{eq:star-H-leads}\\
H_{\mathrm{coupling}}^{\mathrm{(disc)}} & =\sum_{\alpha,\lambda}\underbrace{\sum_{n}v_{\alpha n\lambda}\Bigl(c_{\alpha n}^{\dagger}}_{\equiv t_{\alpha0\lambda}c_{\alpha0}^{\dagger}}d_{\lambda}+\mathrm{H.c.}\Bigr).\label{eq:star-H-coupling}
\end{align}

Following through with this procedure, the dissipators can formally
be derived. If such a path is pursued, the continuum of states of
a specific interval will serve as the environment only of its representative
level, thus resulting in a local dissipator for each level:
\begin{align}
\hspace{-0.1in}\mathcal{\mathcal{D}}_{\alpha n}\rho=\gamma_{n}\left(1-f_{\alpha}\left(\varepsilon_{n}\right)\right) & \left(2c_{\alpha n}\rho c_{\alpha n}^{\dagger}-\left\{ c_{\alpha n}^{\dagger}c_{\alpha n},\rho\right\} \right)\nonumber \\
+\,\gamma_{n}\quad\ f_{\alpha}\left(\varepsilon_{n}\right)\quad\  & \left(2c_{\alpha n}^{\dagger}\rho c_{\alpha n}-\left\{ c_{\alpha n}c_{\alpha n}^{\dagger},\rho\right\} \right),\label{eq:Lindblad-dissipator}
\end{align}
where $\ensuremath{f_{\alpha}\left(\varepsilon\right)\equiv f_{\mathrm{FD}}\left(\varepsilon;\mu_{\alpha},T_{\alpha}\right)}$
is the Fermi-Dirac distribution for lead $\alpha$ (depending on the
lead specific chemical potential and temperature), and $\left\{ \gamma_{n}\right\} $
are referred to as Lindblad driving rates. This structure implies
that when the leads are decoupled from the impurity, i.e., $v_{\alpha n\lambda}=0$,
they are driven to their equilibrium occupation, as expected. The
total Lindblad equation is then given by: 
\begin{equation}
\mathcal{L}\rho=-i\left[H_{\mathrm{total}}^{\mathrm{(disc)}},\rho\right]+\sum_{\alpha n}\mathcal{\mathcal{D}}_{\alpha n}\rho\,,\label{eq:Lindblad-Discrete}
\end{equation}
where $H_{\mathrm{total}}^{\mathrm{(disc)}}\equiv H_{\mathrm{dot}}+H_{\mathrm{leads}}^{\mathrm{(disc)}}+H_{\mathrm{coupling}}^{\mathrm{(disc)}}$
is the total Hamiltonian, now with discrete leads, and hence effectively
of a finite system that becomes an open system by means of the Lindblad
driving. As shown in Ref.~\citep{Schwarz16}, a wide range of driving
rates reproduce the same continuum limit observables. Therefore one
is free to choose them, in this range, so as to best suit the numerics.
With this in mind, and for reasons to be explained in the following
section, the rates will all be chosen to be energy independent, i.e.,
$\gamma_{n}=\gamma$, and of order of the linear level spacing $\delta$.
Let us note that driving of energy modes (exponentially) larger than
$D^{*}$ will have negligible effect on the results, since, importantly,
these modes \emph{start} and practically remain in equilibrium throughout
the dynamics. Thus, the corresponding couplings can further be tuned,
or even completely turned off, in order to enhance numerical stability,
as we further discuss in Sec.~\ref{sec:Error-Analysis} and Appendix
\ref{sec:Appendix-RI-Driving}. At this point, the Lindblad equation
to be solved is fully defined. As a consistency check, note that properly
taking the limits of this equation converges back to the continuum
limit: In the limit $\Lambda\rightarrow1$ the discretization scheme
collapses to a linear (equal spacing) discretization, which in turn
converges to the continuous system in the $\gamma=\delta\rightarrow0$
limit \citep{Schwarz16}.

\subsection{Local Form\label{subsec:Local-Form}}

The Lindblad driven discretized levels (LDDL) scheme \citep{Schwarz16}
is a set of exact manipulations, applied to the Lindblad equation
(\ref{eq:Lindblad-Discrete}) with the goal of bringing it to a form
more favorable for treatment in the framework of tensor-networks.
The system Hamiltonian obtained after discretization is formulated
in the so-called \textit{star} geometry, involving diagonal leads,
as in Eq.~(\ref{eq:star-H-leads}), with all levels directly coupled
to the impurity, as in Eq.~(\ref{eq:star-H-coupling}). This geometry
is non-local (in the sense that \textit{all} lead levels couple to
the impurity), and therefore less convenient in the framework of tensor-networks.
The dissipators, on the other hand, are already local in this geometry,
with each lead level coupled to its own Lindblad bath, which is a
property we would like to retain. A standard procedure, employed for
example in NRG, is to perform an exact mapping in terms of a basis
transformation from the \textit{star} geometry to a \textit{chain}
geometry \citep{Bulla08}. The bilinear structure of the coupling
term in Eq.~(\ref{eq:star-H-coupling}) directly defines the only
bath level $c_{\alpha0}^{\dagger}$ that the impurity couples to.
This level constitutes the first site of a nearest-neighbor tight-binding
chain, which can be obtained by tridiagonalizing the single-particle
basis of the remainder of the lead levels, e.g., by construction of
a full Krylov space:
\begin{equation}
H_{\mathrm{leads}}^{\mathrm{(disc)}}=\sum_{\alpha,k}t_{\alpha k}\left(c_{\alpha k}^{\dagger}c_{\alpha k+1}+c_{\alpha k+1}^{\dagger}c_{\alpha k}\right)+\sum_{\alpha,k}\varepsilon_{\alpha k}c_{\alpha k}^{\dagger}c_{\alpha k}.
\end{equation}
Such a basis transformation, however, will result in non-local dissipators
due to the $n$-dependent prefactors in Eq.~(\ref{eq:Lindblad-dissipator}),
which include the Fermi factors. The LDDL scheme circumvents this
problem and yields a Lindblad equation which is local in both the
dissipators and the Hamiltonian in the \textit{chain} geometry. For
completeness it will be described here briefly. The idea behind this
scheme is to shift the Fermi-Dirac information from the dissipators
in Eq.~(\ref{eq:Lindblad-dissipator}) to the lead-impurity couplings
in an effective Hamiltonian, still in the \textit{star }geometry.
With an appropriate choice of Lindblad driving rates, the system can
then be tridiagonalized into the \textit{chain} geometry, without
loosing the locality of the dissipators.

Consider a single discrete lead level with creation operator $c_{\alpha n}^{\dagger}$,
referred to as a \emph{physical} level, in lead $\alpha$ at energy
$\varepsilon_{n}$ and coupling constants $v_{\alpha n\lambda}$ to
the impurity levels. We temporarily drop the subscripts $\alpha n$
for readability in what follows. Its dissipator is given according
to Eq.~(\ref{eq:Lindblad-dissipator}), meaning it is constantly
depopulated and re-populated at a constant Lindblad driving rate $\gamma$,
weighted by $1-f\left(\varepsilon\right)$ and $f\left(\varepsilon\right)$,
respectively. In the LDDL scheme this single physical level is mapped
onto two artificial lead levels with creation operators $c_{h}^{\dagger}$
and $c_{p}^{\dagger}$, referred to as hole and particle levels, thus
effectively doubling the number of levels. The former is constantly
depopulated at rate $\gamma$ and the latter constantly re-populated
at rate $\gamma$: 
\begin{align}
\mathcal{D}_{h}\rho & =\gamma\left(2c_{h}\rho c_{h}^{\dagger}-\left\{ c_{h}^{\dagger}c_{h},\rho\right\} \right),\label{eq:particle-hole-dissipators}\\
\mathcal{D}_{p}\rho & =\gamma\left(2c_{p}^{\dagger}\rho c_{p}-\left\{ c_{p}c_{p}^{\dagger},\rho\right\} \right).\nonumber 
\end{align}
These two levels both have the same onsite energy $\varepsilon$,
yet are now coupled to the impurity with amplitudes that depend on
temperature and chemical potentials:
\begin{align}
v_{\lambda,h} & =\sqrt{1-f(\varepsilon)}\,v_{\lambda},\qquad v_{\lambda,p}=\sqrt{f(\varepsilon)}\,v_{\lambda}.
\end{align}
Formally this mapping is obtained by introducing an auxiliary level
at energy $\varepsilon$ which is decoupled both from the impurity
and the Lindblad baths, and performing a unitary rotation between
the physical and auxiliary levels, thus shifting the Fermi-Dirac information
from the dissipators to the lead-impurity couplings, as shown in Fig.~\ref{fig:Mappings}(c').
For more details, as well as a discussion of the resemblance of this
procedure to purification of the level, or the thermofield approach,
see Ref.~\citep{Schwarz18}.

The described procedure is repeated for each lead level. This replaces
each physical lead with a corresponding hole lead and particle lead,
as in Fig.~\ref{fig:Mappings}(c), thus doubling the total number
of lead levels. By selecting Lindblad driving rates $\gamma$ to be
energy independent, one obtains dissipators for each of the hole or
particle leads which do not depend on the energy index $n$ (i.e.,
are proportional to to the identity matrix w.r.t.\ this index). Each
such lead can therefore be tridiagonalized separately into a nearest-neighbor
chain while the dissipators remain unaltered, resulting in a Lindblad
equation which is local both in the dissipators and the Hamiltonian,
as desired {[}see Fig.~\ref{fig:Mappings}(d){]}.

Two remarks are in order regarding the doubling of lead levels, before
the tridiagonalization is performed. The first is that for physical
levels lying far from the chemical potential in units of temperature,
where $f\left(\varepsilon\right)$ is 0 (1), the particle (hole) level
decouples from the impurity, and can thus be disregarded in subsequent
calculations. For zero temperature this holds for all physical levels,
and so the described mapping is reduced to relabeling physical levels
above (below) the lead chemical potential as holes (particles), with
no doubling actually occurring.

The second remark relates to exploiting a left-right symmetry in the
lead spectrum. In equilibrium calculations, when both leads have the
same energy levels and for each lead level the left and right coupling
constants to all impurity levels are proportional, only a specific
linear combination of left and right levels couples to the impurity,
precisely as defined by the coupling Hamiltonian. The complementary
orthogonal combination of left and right levels decouples from the
impurity and hence becomes irrelevant for the impurity dynamics. This
simplifies the model from a two-lead model to an effective single-lead
model. In the nonequilibrium case, the different potentials applied
to the left and right leads break this symmetry, and prevent its exploitation.
However, once the physical leads are separated into hole and particle
leads, the symmetry is reinstated (for holes and particles separately),
and can therefore be exploited. Thus for models in which this symmetry
exists, the final number of artificial lead levels is actually smaller
than the original number of physical levels.

\subsection{Renormalized Impurity -- NRG\label{subsec:Renormalized-Impurity}}

The LDDL scheme is indifferent to the specific discretization scheme
employed, as long as the Lindblad driving rates are kept energy independent.
Observe now the implications of the linear-logarithmic scheme on the
resulting Lindblad equation. The obtained on-site energies $\left\{ \varepsilon_{\alpha k}\right\} $
and nearest-neighbor hopping amplitudes $\left\{ t_{\alpha k}\right\} $
in the vicinity of the impurity are of the largest magnitude and decay
exponentially as the distance from the impurity grows, all the way
down to $D^{*}$. The corresponding sites will therefore be referred
to as the \textit{logarithmic} sector. Below $D^{*}$, the on-site
energies and hopping amplitudes remain of order of the linear level
spacing $\delta$ and $D^{*}$, respectively, and will be referred
to as the \textit{linear} sector. Due to the smooth transition in
the discretization, the exact boundaries between these two sectors
are fuzzy, and in practice are chosen with some fine tuning in order
to enhance convergence.

In the chain geometry, the logarithmic sector, including the impurity,
can be considered as a mesoscopic system, coupled to the linear sector
leads. By construction, the vast majority of the (many-body) energy
levels of this mesoscopic system are at energies larger than $D^{*}$,
and so are expected to be indifferent to the voltage bias applied,
thus largely remaining in the equilibrium state. They are therefore
expected to contribute to the nonequilibrium dynamics only through
renormalization effects on the low-energy modes in the linear sector,
which in turn actively participate in the dynamics. As argued in Ref.~\citep{Schwarz18},
it is therefore sufficient to approximate the mesoscopic system by
a renormalized impurity (RI) residing in an effective significantly
reduced low-energy basis. This is a controlled approximation, as one
can monitor the weight on all states in the RI while time-evolving
towards $\rho_{\mathrm{NESS}}.$ Note that the chemical potential
of this RI is set midway between the chemical potentials of the leads,
so that the effective low-energy subspace consists of states with
RI particle number which is close to its average occupation in the
NESS.

The RI is obtained by the following procedure, as described in Fig.~\ref{fig:Mappings}(e'):
An additional subsequent tridiagonalization is applied to merge the
two (particle and hole) chains in the logarithmic sector. This brings
them into a single-lead Wilson-chain structure, which is important
for NRG, since it keeps correlations at a given energy scale local.
An NRG sweep is then applied to the chain -- starting from the impurity,
at each step a site is added to the chain, the Hamiltonian is diagonalized,
and high-energy modes are discarded. At the end of the sweep through
the logarithmic sector, the $R$ lowest-lying many-body states are
taken as the effective basis of the RI. All operators acting on the
impurity, or on sites in the logarithmic sector, are then projected
to this effective reduced basis.

The leads in the linear sector, together with the RI, now form the
dynamical system under consideration, as shown in Fig.~\ref{fig:Mappings}(e).
The Lindblad equation for this system still consists of a nearest-neighbor
Hamiltonian, however with a more complicated local term acting on
the RI site. The dissipators are also local in this setup, and again
the local terms acting on the RI are more complicated, corresponding
to the multiple dissipators acting on the logarithmic sector. Note
that although the dissipators on different sites of the chain originally
commute, the logarithmic sector dissipators, after being projected
to the RI basis, no longer do. Another concern regarding the logarithmic
sector dissipators is that because they were not taken into account
during the RG flow, they might drive the RI out of the effective low-energy
basis. In practice this issue can be handled, as discussed below in
Sec.~\ref{sec:Error-Analysis} and Appendix \ref{sec:Appendix-RI-Driving}

\subsection{MPDO Solution -- tDMRG\label{subsec:MPDO-Solution}}

\begin{figure}[t]
\begin{centering}
\includegraphics[width=1\columnwidth]{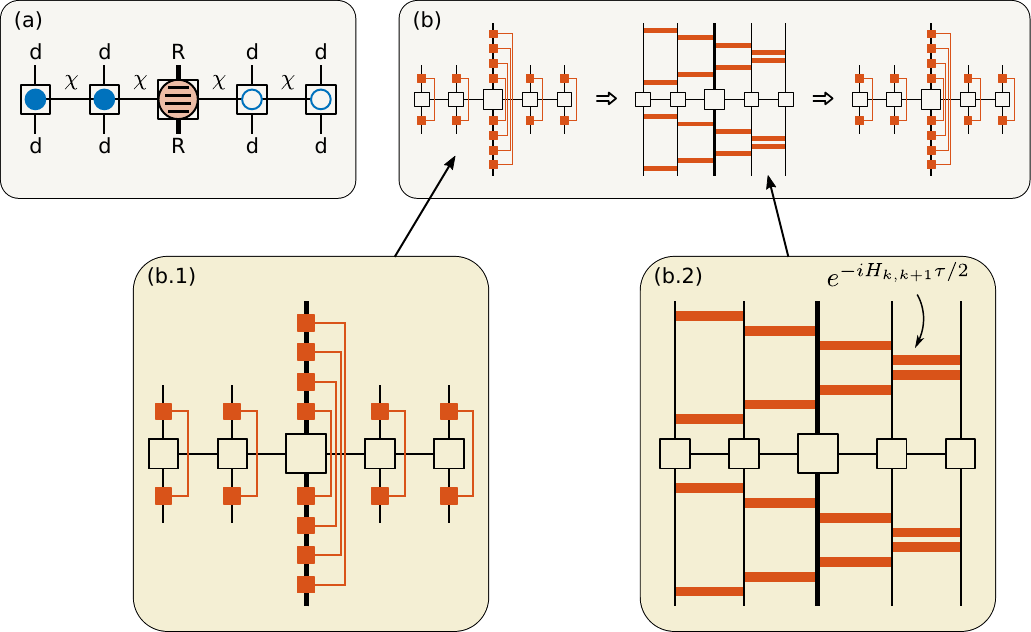}
\par\end{centering}
\caption{(a) MPDO description of the system density operator -- each site
is described by a rank-4 tensor (i.e., 4 legs) with two physical indices,
of local dimension $d$ for chain sites or $R$ for the renormalized
impurity (RI), and two virtual indices, of bond dimension $\chi$,
connecting it to its neighboring tensors. The initial MPDO is chosen
as a product state, where all particle sites are full, all hole sites
are empty, and the RI is in its ground state. (b) A single second-order
Trotterized time-step consists of a dissipative half-time-step sweep
(b.1), a Hamiltonian full time-step sweep (b.2: sweep forward and
backward at $\tau/2$), and then another dissipative half-time-step
sweep in the opposite direction {[}cf.\ Eq.~(\ref{eq:tau-step}){]}.
(b.1) Dissipative evolution half-time-step sweep -- each red square
corresponds to the set of Kraus gates applied to a specific site,
and the sum on all gates at that site is implied by the red contraction
line (see text). At the RI, multiple gates are sequentially applied,
corresponding to the multiple driven sites incorporated into it. (b.2)
Hamiltonian evolution time-step -- each red bar corresponds to a
half-time-step Trotter gate $e^{\pm iH_{k,k+1}\tau/2}$ {[}cf.\ Eq.~(\ref{eq:gate:2site}){]}
applied to sites $k,k+1$. The gates are applied from left to right
and then back form right to left for a full time-step sweep. \label{fig:MPDO}}
\end{figure}
The obtained Lindblad equation is solved for the steady-state by real-time
evolution, implemented in the tensor-network formalism. The system
(mixed) state is represented as a matrix-product density operator
(MPDO) \citep{Verstraete04}. Analogously to the matrix-product state
(MPS) representation of wavefunctions, which has a single (physical)
index for each chain site, the MPDO has two (physical) indices for
each chain site, as shown in Fig.~\ref{fig:MPDO}(a). For the chain
at hand, each of these physical indices is of dimension $d$ corresponding
to a single fermionic Hilbert space, except at the RI, where it is
of dimension $R$ corresponding to the effective low-energy subspace.
It is common practice to combine the two physical indices of each
MPDO site into a single effective index of dimension $d^{2}$ and
simply treat it as an MPS. However, in the case of a large physical
index dimension, e.g., for the RI, keeping the indices separate enables
more efficient contractions and reduces the computation cost.

The derivation of the Lindblad equation respects the same local symmetries
as the original continuous system. In our case the \textit{full} continuous
system conserves $U\left(1\right)$ charge (particle number). The
discretization procedure does not break any of these symmetries, so
that the resulting Hamiltonian still conserves the same charges for
the discrete finite system alone. On the other hand, the derived Lindblad
operators do not necessarily conserve the charge in the finite system
alone, but do respect the symmetry and conserve the charge for the
full system, including the baths. Hence, although the Lindblad dynamics
does not conserve this charge for the finite system, it does conserve
a related quantity \citep{Albert14}, which can be exploited in order
to decompose the MPDO into symmetry sectors, further reducing the
computational cost. We demonstrate this for particle number conservation.
Define the super-operators $\mathcal{N}_{\pm}=N\otimes\mathbb{I}\pm\mathbb{I}\otimes N$,
where $N$ is the particle number operator and $\mathbb{I}$ is the
identity. The Liouvillian super-operator $\mathcal{L}$ commutes with
$\mathcal{N}_{-}$ but not with $\mathcal{N}_{+}$. The dynamics therefore
does not conserve particle number $N\otimes\mathbb{I}=\frac{\mathcal{N}_{+}+\mathcal{N_{-}}}{2}$.
However the conservation of $\mathcal{N}_{-}$ suffices in order to
decompose the MPDO into particle-number symmetry sectors. It also
implies that the parity of $\mathcal{N}_{+}$ and thus of $N\otimes\mathbb{I}$
is conserved, which suffices in order to account locally for fermionic
signs. This example holds for any abelian symmetry, while a more involved
argument is required in the case of non-abelian symmetries.

Following Ref.~\citep{Schwarz18}, the system is set in an initial
state $\k{\psi\left(0\right)}$ which is a product state between the
ground state of the decoupled RI, and the steady-state of the decoupled
linear sector leads. The latter is defined as the pure product state
where all lead particle (hole) sites are full (empty). This initial
state can be written either as an MPS, or as an MPDO for $\rho\left(0\right)\equiv\k{\psi_{0}}\b{\psi_{0}}$,
both with bond dimension $\chi=1$. This starting point is assumed
to be sufficiently close to the desired final steady-state solution,
so that when the coupling to the RI is turned on, the full system
will quickly converge to its steady-state (as our results confirm).
One could also initialize the RI in its decoupled steady-state, but
in practice this does not improve convergence. Note that the initial
setup, together with its transient dynamics, are regarded only as
a means to obtain the desired steady-state, so that the specific choice
of initial state can be fully based on numerical considerations.

The coupling between the RI and the leads is then turned on, and the
system is evolved in time by a variant of tDMRG, formulated to accommodate
for Lindblad dynamics. Note that in this work the RI-lead coupling
is turned on in an immediate quench, and slow ramping up of the coupling,
as employed in Ref.~\citep{Schwarz18}, was not necessary. In the
spirit of tDMRG, this time-evolution is based on a second-order Trotter-Suzuki
decomposition with a sufficiently small time-step $\tau$. Then the
propagator can be written as a product of short-time propagators $e^{\mathcal{L}t}=\prod_{i=1}^{N_{t}}e^{\mathcal{L}\tau}$,
with $N_{t}$ steps required in order to arrive at a time $t=\tau N_{t}$.
Each short-time propagator is Trotter decomposed into local and nearest-neighbor
gates based on the short-rangedness of the Liouvillian introduced
above. The total Hamiltonian can be written as a sum of local two-site
operators $H=\sum_{k=1}^{N-1}H_{k,k+1}$ where non-adjacent terms
commute. Defining the Hamiltonian two-site super-operators as $\mathcal{H}_{k,k+1}\rho\equiv-i\left[H_{k,k+1},\rho\right]$,
the Liouvillian $\mathcal{L}$ can then be written as the sum of these
two-site Hamiltonian terms and single-site dissipative (hole/particle)
terms defined in Eq.~(\ref{eq:particle-hole-dissipators}): 
\begin{equation}
\mathcal{L}=\sum_{k=1}^{N-1}\mathcal{H}_{k,k+1}+\sum_{k=1}^{N}\mathcal{D}_{k}.
\end{equation}
For an exact representation of the super-operators, the Hamiltonian
terms commute with all non-adjacent Hamiltonian and dissipative terms,
and the dissipative terms all commute with each other. However, inside
the RI the fermionic anti-commutation relations of the original fermionic
operators are compromised by the NRG truncation, which results in
non-commuting terms in its vicinity. The second-order Trotter decomposition
adopted here and depicted in Fig.~\ref{fig:MPDO}(b), is similar
to the one discussed in Ref.~\citep{Werner16}:
\begin{equation}
e^{\mathcal{L}\tau}\approx\underbrace{\prod_{k=1}^{N}e^{\frac{\tau}{2}\mathcal{D}_{k}}}_{\text{Fig.\,\ref{fig:MPDO}(b.1)}}\cdot\underbrace{\prod_{k=N}^{2}e^{\frac{\tau}{2}\mathcal{H}_{k-1,k}}\cdot\prod_{k=1}^{N-1}e^{\frac{\tau}{2}\mathcal{H}_{k,k+1}}}_{\text{Fig.\,\ref{fig:MPDO}(b.2)}}\cdot\underbrace{\prod_{k=N}^{1}e^{\frac{\tau}{2}\mathcal{D}_{k}}}_{\text{Fig.\,\ref{fig:MPDO}(b.1)}}.\label{eq:tau-step}
\end{equation}
The two-site Hamiltonian gates are given by: 
\begin{equation}
e^{\frac{\tau}{2}\mathcal{H}_{k,k+1}}\rho\equiv e^{-i\frac{\tau}{2}H_{k,k+1}}\rho\,e^{i\frac{\tau}{2}H_{k,k+1}}.\label{eq:gate:2site}
\end{equation}
and the dissipative single-site gates translate into Kraus operators
\citep{KRAUS1971311,nielsen_chuang_2010}. In the spinless case they
are respectively given for particles or holes by: 
\begin{align}
 & e^{\frac{\tau}{2}\mathcal{D}_{\eta}}\rho=K_{1\eta}\rho K_{1\eta}^{\dagger}+K_{2\eta}\rho K_{2\eta}^{\dagger}\qquad\eta\in\left\{ h,p\right\} ,\label{eq:Kraus}\\
 & K_{1h}=e^{-\frac{\tau}{2}\gamma c^{\dagger}c}=cc^{\dagger}+e^{-\gamma\tau/2}c^{\dagger}c\ \quad K_{2h}=\sqrt{1-e^{-\gamma\tau}}c,\nonumber \\
 & K_{1p}=e^{-\frac{\tau}{2}\gamma cc^{\dagger}}=c^{\dagger}c+e^{-\gamma\tau/2}cc^{\dagger}\ \quad K_{2p}=\sqrt{1-e^{-\gamma\tau}}c^{\dagger}.\nonumber 
\end{align}
For spinfull fermions there will be 4 Kraus operators for each site,
replacing $\eta\to\left(\eta,\sigma\right)$, with $\sigma\in\left\{ \uparrow,\downarrow\right\} $.
For terms which are bilinear in the fermionic operators, such as the
Hamiltonian or $K_{1\eta}$, fermionic signs arising from the anti-commutation
relations can be accounted for locally. The operators $K_{2\eta}$
in the dissipative terms, however, act simultaneously on both sides
of the density matrix. Hence, they give rise to a global Jordan-Wigner
string. In the present MPDO setup, it can be efficiently `pulled'
in locally \citep{Corboz10}. This requires that charge parity is
fully tracked on all tensors, which is the case here when decomposing
the MPDO into $U\left(1\right)$ charge symmetry sectors, in the sense
discussed above. In the local configuration, as shown in Fig.~\ref{fig:MPDO}(b.1),
the crossing of the red line with the bond index implies that the
charge parity operator $Z\equiv\left(-1\right)^{q}$, with charge
$q$, must be simultaneously applied to the bond state space when
acting with $K_{2\eta}$.

A quick overall complexity analysis of the method can be performed
assuming a fixed bond dimension $\chi$ on all MPDO sites. Since the
treatment of the RI is clearly the most expensive step, the following
considers operations involving the RI. The analysis is completely
analogous for all other sites where one just replaces $R$ with the
regular local dimension $d$ of a physical site. The cost of the Trotter
gate contraction is $O\left(d^{2}R^{2}\chi^{3}+d^{3}R^{3}\chi^{2}\right)$,
where the two terms correspond to merging the RI tensor with its neighboring
tensor and to applying the nearest-neighbor Trotter gate, respectively.
The SVD back into local tensors then costs $O\left(d^{4}R^{2}\chi^{3}\right)$.
Finally, the cost of the Kraus gates contractions is $O\left(kR^{3}\chi^{2}\right)$,
where $k$ is the number of Kraus gates acting on the RI, which is
proportional to the number of sites in the logarithmic\textit{ }sector.
Ignoring the cost of all other sites in the linear sector, the total
cost of the method can be approximated as $O\left(N_{t}\left(d^{3}R+kR+d^{4}\chi\right)R^{2}\chi^{2}\right)$,
where $N_{t}$ is the number of sweeps.

The most important property of the MPDO ansatz is that it can efficiently
represent the steady-state, using a relatively small number of parameters.
Another important constraint on the represented state is that it must
be a physical state, i.e., a positive semi-definite Hermitian operator
with finite trace. While the MPDO ansatz does not enforce these constraints,
the Lindblad evolution (also after Trotterization) is a completely-positive
trace-preserving (CPT) map \citep{nielsen_chuang_2010}, and thus
guarantees that starting from a physical state will always result
in a physical state. The only loss of positivity (and trace) can come
from truncation of singular values during the tDMRG sweep. This relates
to a drawback of the MPDO ansatz -- the singular values obtained
after a Schmidt decomposition no longer correspond to the singular
values of the reduced density matrix (as for an MPS). However, if
the singular values drop quickly enough, as is the case for the models
analyzed here, only small singular values are truncated, resulting
in negligible loss of positivity.

\subsection{Observables\label{subsec:Observables}}

At any point throughout the evolution, single-time correlations can
be extracted from the MPDO, with the long-time limit representing
the steady-state value. As correlations in the chain geometry are
easily obtained, in practice it is convenient to map all observables
of interest to this geometry, as shown in Ref.~\citep{Schwarz18}.
In this work we focus on the particle current flowing from one lead
to the other. The time derivative of the impurity occupation can be
separated into contributions $I_{\alpha}$ corresponding to the current
flowing from lead $\alpha$ into the impurity:
\begin{equation}
-e\tfrac{d}{dt}\langle n_{d}\rangle=-\tfrac{ei}{\hbar}\left\langle \left[H,n_{d}\right]\right\rangle =\sum_{\alpha}I_{\alpha}.
\end{equation}
 Thus $I_{\alpha}$ is given in the continuum limit (taking $e=1,\ \hbar=1$),
and approximated after discretization by:
\begin{equation}
I_{\alpha}=2\sum_{\lambda}v_{\alpha\lambda}\int_{-D}^{D}\tfrac{d\varepsilon}{\sqrt{2D}}\mathrm{Im}\left\langle c_{\alpha\varepsilon}^{\dagger}d_{\lambda}\right\rangle \approx2\sum_{n,\lambda}v_{\alpha n\lambda}\mathrm{Im}\left\langle c_{\alpha n}^{\dagger}d_{\lambda}\right\rangle .
\end{equation}
 In the steady-state, $\tfrac{d}{dt}\langle n_{d}\rangle=0$, the
current flowing from the left lead into the impurity is equal to the
current flowing from the impurity into the right lead $I\equiv I_{L}=-I_{R}$.
From a numerical perspective, the average combination $I=\left(I_{L}-I_{R}\right)/2$
converges more rapidly, and is less prone to noise. Running simulations
for different voltages, a full $I$-$V$ curve can be obtained, and
numerically differentiated in order to produce the differential conductance
$G\left(V\right)=\Delta I/\Delta V$. Note that the numerical derivative
is very sensitive to noise, so that the $I$-$V$ curve must be obtained
with high accuracy.

\section{Error Analysis and the RLM\label{sec:Error-Analysis}}

\begin{figure*}[t]
\begin{centering}
\includegraphics[width=1\textwidth]{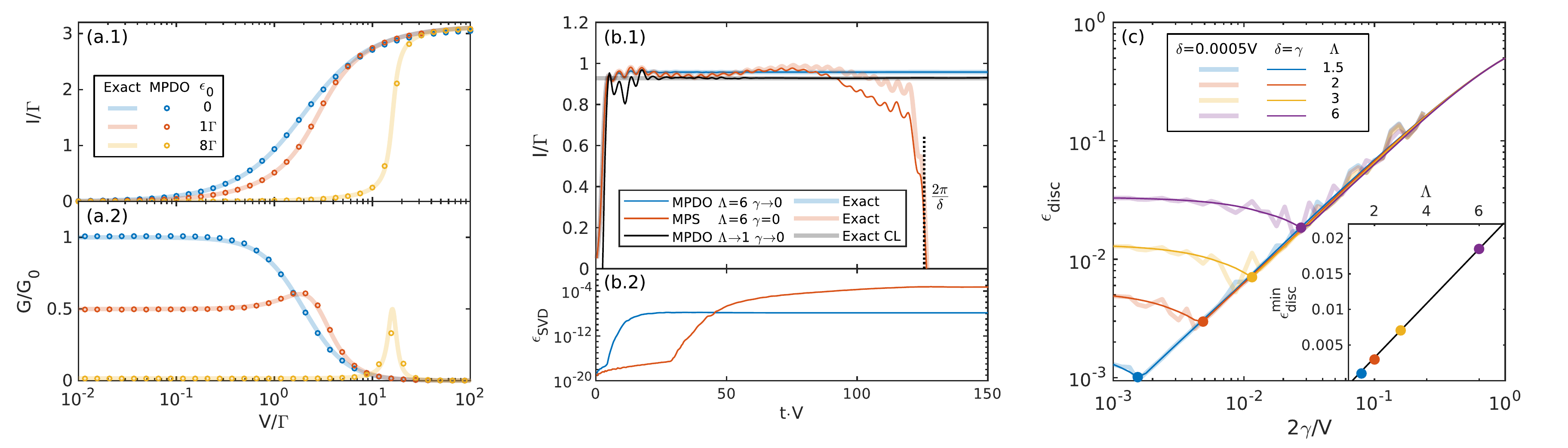}
\par\end{centering}
\caption{(a.1) NESS current and (a.2) differential conductance of the RLM (in
units of the conductance quantum $G_{0}=e^{2}/h$) as a function of
bias voltage $V$ and at different gate voltages $\varepsilon_{0}$,
as calculated exactly (solid lines) and by RL-NESS (circles). Simulation
results are obtained for $\delta=0.1D^{*},\ D^{*}=\frac{V}{2}$ and
after linear extrapolation in $\gamma/\delta=2,1\rightarrow0$ and
$\Lambda=8,6\rightarrow1$, with bond dimension $\chi=256$ and $R=32$
states kept in the RI. (b.1) Comparison of the RL-NESS MPDO evolution
($\chi=256$), after extrapolation to $\gamma\rightarrow0$ (blue),
and the NRG-tDMRG MPS evolution ($\chi=1024$) for $\gamma=0$ (red),
both with the same discretization, at bias voltage $V=\Gamma$ and
$\varepsilon_{0}=0$. Exact results for both cases are also plotted
(shaded). The MPDO result is then further linearly extrapolated from
$\Lambda=8,6$ to $\Lambda=1$ (solid black) and compared to the exact
continuous leads (CL) result (shaded gray). (b.2) Truncation error
(sum of singular values squared accumulated over several time-steps)
plotted for the RL-NESS MPDO (blue) and NRG-tDMRG MPS (red) evolution.
(c) Discretization error $\epsilon_{\mathrm{disc}}$ for the RLM at
$\varepsilon_{0}=0$ as a function of $\gamma$ for different values
of $\Lambda$, with $\delta=0.0005V$ (solid) and $\delta=\gamma$
(shaded). The inset displays the $\Lambda$-dependence of the lower
bound on the discretization error $\epsilon_{\mathrm{disc}}$ (corresponding
to the circles in the main figure). \label{fig:RLM-Error}}
\end{figure*}

In order to estimate the accuracy of the presented method, we apply
it to the exactly solvable non-interacting resonant level model (RLM).
This model, with dot Hamiltonian 
\begin{equation}
H_{\mathrm{dot}}^{RLM}=\varepsilon_{0}n_{0},
\end{equation}
describes a single spinless impurity level with energy $\varepsilon_{0}$
(e.g., controlled by a gate voltage), coupled to two spinless leads
described by Eq.~(\ref{eq:CL-H-coupling}) via the coupling Hamiltonian
(\ref{eq:CL-H-leads}). An end-to-end comparison of the steady-state
current and the differential conductance, between the exact result
of the RLM in the continuum limit and the RL-NESS result, is shown
in Fig.~\ref{fig:RLM-Error}(a). It displays a good agreement over
a wide range of bias voltages and impurity level energies $\varepsilon_{0}$,
with parameter values given in the caption.

The RL-NESS real-time evolution of the current, for a typical case
of $V=\Gamma,\ \varepsilon_{0}=0$ with finite $\Lambda=6$ and linearly
extrapolated to $\gamma\rightarrow0$, is plotted in Fig~\ref{fig:RLM-Error}(b)
(blue). It demonstrates several key aspects of the method. After an
initial rapid rise in the current over a period of order $V^{-1}$,
the oscillations (discretization artifacts related to the logarithmic
sector) decay exponentially at a rate which is proportional to $\gamma$,
finally stabilizing on a steady-state value. For further discussion
regarding the evolution time scales see Appendix \ref{sec:Appendix-Evolution-Time-Scales}.
The convergence to the steady-state can also be observed in the lower
panel, where the truncation error saturates. Throughout the entire
evolution, our method displays excellent agreement with the corresponding
exact result (shaded blue) of the same driving protocol. After linearly
extrapolating also to the $\Lambda\rightarrow1$ limit, the RL-NESS
current (black), once the NESS is reached, displays excellent agreement
with the continuum limit exact result (shaded gray).

If the dissipation is initially set to $\gamma=0$, while keeping
a finite level spacing, RL-NESS reduces to the NRG-tDMRG scheme \citep{Schwarz18},
in which the state of the system is represented by an MPS (instead
of an MPDO), and the real-time evolution is unitary. In what follows
this will simply be referred to as MPS evolution. As in the case of
finite dissipation, the $\gamma=0$ evolution of the current can be
calculated (for the same $\Lambda=6$) either explicitly as an MPS
evolution (solid red) or exactly in the single particle basis (shaded
red). These results agree with the RL-NESS current in the early transient
oscillatory regime, but later residual oscillations persist. Thus
only a quasi-steady-state is obtained, whose mean is nevertheless
consistent with the $\gamma\rightarrow0$ limit. Eventually, the current
drifts away, due to truncation errors that, without dissipation, do
not saturate. Later on, even for the exact solution, this quasi-steady-state
will be lost due to reflection off the edges of this closed system
at a time $t=L/v_{F}=\frac{2\pi}{\delta}$, dictated by the finite
linear level spacing. In stark contrast, in the case of RL-NESS, for
strong enough damping $\gamma\apprge\delta$, the discrete levels
become sufficiently blurred out, such that the dynamics truly represents
an open system, where reflection off the edges and the accompanying
drop in the current no longer occur.

As part of the analysis, steady-state observables of the RLM are calculated
exactly by means of Keldysh formalism, both for the continuous system
and in an arbitrary discretization (Appendix \ref{sec:Appendix-Exact-Solution}).
The exact time evolution of single-time observables (in a given discretization)
is also calculated by solving a differential continuous Lyapunov equation
for the single-particle correlation matrix (Appendix \ref{sec:Appendix-Lyapunuv-Equation}).

The remainder of this section is dedicated to an analysis of the two
major error contributions in the method: the lead \textit{discretization
error} -- how well do the discrete system observables represent the
continuous system, and the \textit{simulation error} -- how accurately
does the tensor-network method solve for the discrete system steady-state.
Generally there is a trade-off between the two contributions, as a
finer discretization better reproduces the continuum limit, but is
also harder to solve for numerically.

The lead discretization error depends both on the fineness of the
discretization grid, controlled by the logarithmic $\Lambda$ and
linear $\delta$ discretization parameters, and on the broadening
of the discrete levels, controlled by the Lindblad driving rates $\gamma$.
We introduce the relative measure for the discretization error, $\epsilon_{\mathrm{disc}}\equiv\max_{V}\left|1-I_{DL}/I_{CL}\right|$,
as the maximal relative distance over a range of bias voltages $V\in\left[0.01,100\right]\Gamma$,
between the exact discrete leads current $I_{DL}$ (for a specific
choice of $\Lambda$ and ratios $\delta/V,\gamma/V$) and the exact
continuous leads current $I_{CL}$. This measure can be explicitly
evaluated for the RLM and is plotted in Fig.~\ref{fig:RLM-Error}(c)
as a function of $\gamma/V$ and for several values of $\Lambda$.
The specific choice of $\delta$ has only a minor effect, as long
as $\delta\leq\gamma$, which is required in order to negate finite
size effects. Fixing $\delta/V$ to a small value results in a smooth
curve (solid), while taking $\delta=\gamma$ results in a slightly
more noisy curve (shaded) with the same trend. Note that $\epsilon_{\mathrm{disc}}$
is approximately linear in $\gamma/V$, down to a lower bound on the
error, dictated by $\Lambda$. This lower bound in turn is linear
in $\Lambda$ {[}see inset to Fig.~\ref{fig:RLM-Error}(c){]}. These
two observation justify a linear extrapolation in these two parameters
to the continuum limit $\Lambda\rightarrow1,\ \delta=\gamma\rightarrow0$
at each $V$.

The simulation error has multiple contributions, listed in ascending
order of significance. First consider the Trotter error, arising from
the discretization of the Liouvillian real-time evolution. In practice,
exploiting second-order Trotter decomposition, it is numerically feasible
to choose sufficiently small time-steps such that this error is negligible
compared to the other ones. A second source of simulation error is
introduced by the NRG procedure, and controlled by the number of kept
states in each NRG iteration. As in equilibrium, the number of required
kept states can be reduced by taking a coarser logarithmic discretization,
i.e., larger $\Lambda$. In practice, only the number of kept states
$R$ in the last NRG iteration, dictating the size of the restricted
low-energy subspace of the RI, pose a computational bottleneck. The
numerical cost in setting up the RI by previous NRG iterations is
entirely negligible. Therefore earlier NRG iterations are less harshly
truncated, but a larger $\Lambda$ is still required in any case in
order to keep $R$ sufficiently small. The third, and most significant,
source of simulation error is the truncation of the MPDO to a fixed
bond dimension $\chi$ after each time-step, by discarding small singular
values. Empirically the singular values decay faster than polynomially
with the singular value index, at a rate which decreases with decreasing
$\gamma$ (see Appendix \ref{sec:MPDO-Singular-Values}). This implies
that the required bond dimension (for a fixed truncation error) scales
exponentially with $\gamma$. This exponential scaling can naturally
be understood in the $\gamma\rightarrow0$ limit, in which the entanglement
entropy grows linearly in time, thus leading to an exponential blowup
in the required bond dimension. Choosing a finite $\gamma$ sets a
time scale $1/\gamma$ at which the entanglement entropy stops growing.
For any finite $\gamma$ the steady-state can therefore be represented
with a finite (possibly large) bond dimension, which in the small
$\gamma$ limit must grow exponentially with $1/\gamma$ in order
to match the expected exponential blowup. It is important to stress
that even though there is an exponential bound on simulating small
$\gamma$, this represents the thermodynamic limit, which can be approached
by working with finite $\gamma$ and then linearly extrapolating to
small $\gamma$.

Finally, let us discuss the issue of whether or not to apply the Lindblad
terms coupled to the RI. Physically, since the RI represents the degrees
of freedom far above the voltage and temperature bias scales, it is
reasonable to expect that they are barely affected by the nonequilibrium
conditions. Thus whether or not the Lindblad terms acting on the RI
are applied, we expect to obtain similar results. We demonstrate this
for the RLM in Appendix \ref{sec:Appendix-RI-Driving}. Numerically,
however, the effort involved in the two approaches (for the same accuracy)
is different. The effect on the numerical results becomes more pronounced
in the interacting case, considered in the next section. There, for
a large logarithmic discretization parameter $\Lambda$, the RI spectrum
contains nearly degenerate levels, which can be coupled even by weak
Lindblad driving at the sites composing the RI. In practice this can
drive, and hence affect, high-energy modes in the RI (which in principle
should remain in equilibrium), resulting in artifacts which are enhanced
in the differential conductance. Taking small values of $\Lambda\sim2$
could resolve this problem. However, this necessitates increasing
the number $R$ of states kept in the RI, and therefore is often impractical.
Taking the manageable intermediate value $\Lambda=3$ for the interacting
case (instead of extrapolating to $\Lambda\rightarrow1$), at the
cost of a larger $R=64$, suppresses these artifacts, but still does
not completely eliminate them. For these reasons, in the interacting
case it becomes preferable to entirely turn off the Lindblad terms
coupled to the RI. This does not adversely affect the physics. To
the contrary, it leads to a stable numerical solution without artifacts,
with reasonable computational costs.

\section{Interacting System\label{sec:Interacting-System}}

\begin{figure*}[t]
\begin{centering}
\includegraphics[width=1\textwidth]{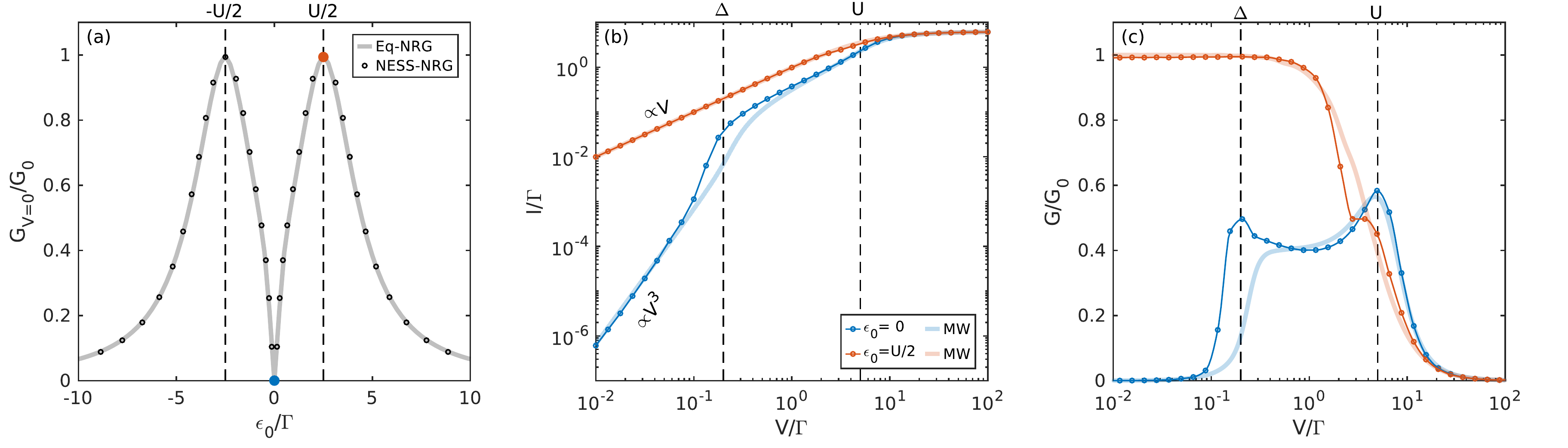}
\par\end{centering}
\caption{RL-NESS simulation results for the interacting two-level model (I2LM)
with level spacing $\Delta=\frac{\Gamma}{5}$ and interaction $U=5\Gamma$,
marked by dashed lines. The simulation is run with discretization
parameters $\Lambda=3,\ \delta=0.1D^{*},\ D^{*}=\frac{V}{2}$, linearly
extrapolated to $\gamma\rightarrow0$ from $\gamma/\delta=4,2,1$,
and with simulation parameters $\chi=256,\ R=64$. (a) Zero bias linear
conductance as a function of gate voltage $\varepsilon_{0}$, calculated
by RL-NESS at $V=0.01\Gamma$ (circles) and compared to the numerically
exact equilibrium NRG result (shaded). (b) NESS current and (c) its
derived differential conductance at finite bias, as calculated by
RL-NESS for two gate voltages, corresponding to the valley at $\varepsilon_{0}=0$
(blue) and the peak at $\varepsilon_{0}=\frac{U}{2}$ (red). The low-bias
behavior of the current exhibits a linear dependence for the peak,
but a cubic dependence for the valley, where the linear response conductance
thus vanishes quadratically in $V$. For comparison the equilibrium
spectral function of the I2LM is plugged into the Meir-Wingreen formula
for the current and conductance (shaded). Note that it quantitatively
captures the small- and large-bias features, but qualitatively misses
various physical features in the intermediate bias regime.\label{fig:I2LM}}
\end{figure*}

We now wish to demonstrate the method on an interacting system, which
has no known solution for the NESS current. For this we choose an
interacting two-level model (I2LM), consisting of two interacting
dot levels $\varepsilon_{1},\varepsilon_{2}$ with onsite interaction
energy $U$, coupled to non-interacting leads. The dot levels are
taken with level spacing $\Delta\equiv\varepsilon_{2}-\varepsilon_{1}$,
and can be shifted by changing $\varepsilon_{0}$ by a gate voltage
(taken relative to particle-hole symmetry), such that the dot Hamiltonian
is given by:
\begin{align}
H_{\mathrm{dot}}^{I2LM} & =\varepsilon_{1}n_{1}+\varepsilon_{2}n_{2}+Un_{1}n_{2},\label{eq:I2LM-H}\\
\varepsilon_{1,2} & \equiv\varepsilon_{0}-\tfrac{1}{2}U\mp\tfrac{1}{2}\Delta.\nonumber 
\end{align}
Both levels are coupled symmetrically to the left and right leads,
so that the lead and coupling Hamiltonians are given by Eqs.~(\ref{eq:CL-H-leads})
and (\ref{eq:CL-H-coupling}), with equal hybridization $\Gamma_{\alpha\lambda}=\Gamma$.
For simplicity, we take all the dot-lead couplings to be real and
with the same sign. Since our main goal is demonstrating the method
rather than studying the model, we do not explore the full impurity
parameter space, but concentrate on restricted yet representative
sets of parameter values. The level spacing and the interaction are
fixed to $\tfrac{U}{5}=\Gamma=5\Delta$ such that $\Delta<\Gamma<U$,
thus having a separation of energy scales in a strongly correlated
regime, and the bias and gate voltages are varied.

First we explore the zero-bias linear conductance for small bias voltage.
Due to the Fermi liquid nature of the low-energy fixed point, the
$T=0$ linear response conductance is determined by the total phase
shift, which in turn is set by the Friedel sum rule \citep{Goldstein_2007,Karrasch_2007},
leading to the relation:
\begin{equation}
G=G_{0}\sin^{2}\left(\pi n_{d}\right),
\end{equation}
with $G_{0}=\frac{e^{2}}{h}$ the conductance quantum, and $n_{d}$
the total dot occupation in equilibrium, which can be calculated by
equilibrium NRG. We show our results for $V=0.01\Gamma\ll\Delta,\Gamma,U$
in Fig.~\ref{fig:I2LM}(a) (circles) vs.\ NRG (shaded). At gate
voltage $\varepsilon_{0}=0$ the system is particle-hole symmetric,
and the impurity is occupied exactly by one electron, such that the
linear conductance vanishes. At $\varepsilon_{0}\approx\mp\frac{U}{2}$
the dot population is close to $1\pm\frac{1}{2}$ respectively, hence
the linear conductance features Coulomb blockade peaks with height
$G_{0}$ (red dot).

Next the NESS current at finite bias is shown in Fig.~\ref{fig:I2LM}(b)
for two gate voltages $\varepsilon_{0}=0$ and $\frac{U}{2}$, corresponding
to the valley and the peak (in the zero-bias conductance). For $\varepsilon_{0}=\frac{U}{2}$
the low-bias behavior exhibits a linear dependence (shaded red), as
expected. For $\varepsilon_{0}=0$ however, the linear response term
vanishes by symmetry, and as the current is an odd function of the
bias voltage, the next term is expected to be cubic in the bias voltage,
as is indeed observed (shaded blue). In the limiting regime of very
large bias $V$, exceeding all other energy scales (except for bandwidth
$D$), the current saturates for both cases to the maximal value of
$2\pi\Gamma$, directly corresponding to the two conduction channels
at coupling strength $\Gamma$ each. The differential conductance
is shown in Fig.~\ref{fig:I2LM}(c), with peaks corresponding to
conductance channels opening up. For $\varepsilon_{0}=0$ we get very
clear peaks, with the first conductance channel opening at $\Delta$
with sequential tunneling and thus fluctuations between the two dot
levels, and the second conductance channel opening at $U$, corresponding
to full charge fluctuations in the dot occupation. For $\varepsilon_{0}=\frac{U}{2}$,
a single-particle level is midway between the two chemical potentials,
so there is already a single channel fully open at zero-bias, resulting
in a differential conductance of $G_{0}$. The conductance starts
dropping close to $V=\Gamma$ to about half its value. In the vicinity
of $V=U$ there is a shoulder, beyond which the conductance drops
to zero since the current saturates.

As an interesting comparison consider an approximate form of the Meir-Wingreen
formula for the-steady state current \citep{Meir92}. The exact version
of this formula reads for the I2LM
\begin{equation}
I=\frac{i}{2}\int_{-\frac{V}{2}}^{\frac{V}{2}}d\omega\mathrm{tr}\left\{ \Gamma\left(\omega\right)\left(G^{R}\left(\omega\right)-G^{A}\left(\omega\right)\right)\right\} ,\label{eq:MW-I2LM-Current}
\end{equation}
where $G^{R,A}\left(\omega\right)$ are the exact retarded and advanced
impurity nonequilibrium Green's functions ($2\times2$ matrices for
the 2 impurity modes), and $\Gamma\left(\omega\right)=\Gamma\left(\begin{array}{cc}
1 & 1\\
1 & 1
\end{array}\right)$, corresponding to symmetric and equal hybridization of both modes
to the two leads (for details see Appendix \ref{sec:Appendix-Exact-Solution}).
There is of course no known expression for the nonequilibrium Green's
functions. However, one could calculate the equilibrium ($V=0$) spectral
function, e.g., by fdm-NRG \citep{Wb07}, and plug it into Eq.~(\ref{eq:MW-I2LM-Current}).
This approximation is valid in the linear response regime, and is
expected to produce quantitatively good results for large bias, but
in the intermediate regime is uncontrolled. Fig.~\ref{fig:I2LM}
therefore also shows the steady-state current (b) and differential
conductance (c) obtained in this manner (shaded).

We see the the equilibrium spectral function results agree quantitatively
with our nonequilibrium results in the low- and large-bias limits.
Interestingly, for $\varepsilon_{0}=0$ they also capture the charge-fluctuation
peak at $V=U$ in the intermediate region, but only hint at the level
fluctuation peak at $V=\Delta$. On the other hand, for $\varepsilon_{0}=\frac{U}{2}$,
they completely miss the shoulder in the drop of the conductance.
Thus we conclude that the RL-NESS method successfully reproduces the
current and conductance in the known limits, but also gives numerically
convergent results in the intermediate regime.

\section{Discussion\label{sec:Discussion}}

To conclude, in this work we have introduced RL-NESS, a new numerically
exact algorithm for finding the steady-state of general impurities
far from equilibrium. It builds on the power of equilibrium NRG in
addressing equilibrium quantum impurities with widely-separated bare
and emergent energy scales, and brings it into the nonequilibrium
realm. The method is based on coherently coupling the impurity to
appropriately log-linearly discretized leads, which in turn are subject
to weak Lindblad driving representing incoherent reservoirs. This
model setup corresponds to the physical picture of, e.g., a quantum
dot coherently coupled to quantum wires, which are in turn coupled
to a classical voltage bias source. The resulting system is numerically
simulated by a combination of NRG reduction of the high-energy degrees
of freedom, followed by tDMRG-based MPDO Lindblad evolution. We benchmark
our approach by presenting results for both noninteracting and interacting
models. The accuracy of these demonstrate the power of our method,
accompanied with a detailed analysis of all error sources and their
treatment.

One can envision different ways to try to improve the algorithm. Having
shown that an efficient representation of the steady-state as a tensor-network
exists, it would be useful to search for more compact representations.
One candidate for such a representation is the locally purified tensor-network
(LPTN) ansatz \citep{Verstraete04,Werner16}, which enforces physical
constraints on the density operator such as positivity, and as such
resides in a smaller manifold. However it is not guaranteed that such
an ansatz will efficiently capture the entanglement structure of the
steady-state \citep{Cuevas_2013}, as preliminary investigation seems
to suggest for the case at hand. So-called disentanglement schemes
for the ancilla index \citep{Hauschild18} might improve the situation,
but require further investigation. Recent works \citep{krumnow19,Rams20}
claim that the entanglement structure of the chain geometry is not
optimal, suggesting that applying time evolution in the star geometry
might result in a slower growth of entanglement entropy, thus requiring
a smaller bond dimension. Testing this idea together with RL-NESS
is left for future work. Another direction which might be worth investigating
is directly solving the Lindblad equation $\mathcal{L}\rho=0$ for
the steady-state \citep{Cui15,Mascarenhas15}, instead of obtaining
it by real-time evolution.

It would be interesting to apply RL-NESS to more complicated models,
such as the single impurity Anderson model \citep{hewson_1993}, the
interacting resonant level model \citep{gogolin2004bosonization},
and the I2LM with non-symmetric coupling \citep{Karrasch07PRL,Goldstein10},
all of which are expected to demonstrate Kondo-like physics. RL-NESS
already incorporates a temperature for each lead, and so can immediately
be employed for finite temperature calculations, as well as calculating
thermal conductance, by assigning a different temperature to each
lead. We also plan to leverage the success of RL-NESS in obtaining
a stable steady-state solution, in order to extract dynamical properties,
i.e., time correlators and spectral functions. In the longer run,
we envision the treatment of far-from-equilibrium higher-dimensional
correlated quantum systems, using, e.g., the dynamical mean field
approach \citep{Sakai94,Bulla98,Stadler15}.
\begin{acknowledgments}
We thank M. C. Bañuls, J. I. Cirac, J. Eisert, F. Schwarz and A. Werner
for helpful discussions. ML thanks F. Schwarz for making her code,
developed for Ref.~\citep{Schwarz18}, available to him at the initial
stages of this project. This joint work was supported by the German
Israeli Foundation (Grant No.\ I-1259-303.10). In addition, AW was
supported by the U.S. Department of Energy, Office of Basic Energy
Sciences (Contract No.\ DE-SC0012704). JvD was supported by the Deutsche
Forschungsgemeinschaft under Germany's Excellence Strategy--EXC-2111--390814868.
MG acknowledges additional support by the Israel Science Foundation
(Grant No.\ 227/15), the US-Israel Binational Science Foundation
(Grant No.\ 2016224), and the Israel Ministry of Science and Technology
(Contract No. 3-12419).
\end{acknowledgments}

\appendix
\setcounter{equation}{0}
\renewcommand{\theequation}{\thesection.\arabic{equation}}

\section{Linear-Logarithmic Discretization\label{sec:Appendix-Lin-Log-Disc}}

As specified in Sec.~\ref{subsec:Lindblad-Equation}, the intervals
$I_{n}$ are chosen such that in the range $\left[-D^{*},+D^{*}\right]$
they are of size $\delta$ and far from this range they scale exponentially
as $\sim\Lambda^{n}$. This choice of intervals is achieved by defining
a function $f\left(x\right)$ for positive $x$, which is linear for
$x<\frac{D^{*}}{\delta}$ and has a smooth transition to exponential
$\sim\Lambda^{x}$ for large $\left|x\right|$:

\begin{align}
I_{n\geq0} & =\left[f\text{\ensuremath{\left(n+z\right)}},f\left(n+1+z\right)\right],\\
f\left(x\right) & =\begin{cases}
\tfrac{\delta}{\log\Lambda}\sinh\bigl((x-\tfrac{D^{*}}{\delta})\log\Lambda\bigr)+D^{*} & x>\tfrac{D^{*}}{\delta}\\
\qquad\qquad\delta\cdot x & x<\tfrac{D^{*}}{\delta}
\end{cases},
\end{align}
with $n$ running on all integers such that the full band is covered
up to the cutoff $D$. The edge of the last interval is then manually
fixed to be $D$. The parameter $z\in\left[0,1\right)$ is referred
to as the $z$-shift parameter (in the NRG jargon), and can be used
to shift the lead energy levels. Since different $z$-shifts result
in different yet equivalent discretizations, it is common practice
to average simulations using different $z$-shifts in order to reduce
numerical artifacts due to the discretization \citep{Oliveira94},
especially when calculating spectral functions. However in this work
it was sufficient to use $z=0$. The intervals for negative energies
are taken as a mirror image of the positive intervals. This guarantees
particle-hole symmetry for any $z$, at the cost of the interval closest
to 0 not necessarily being of size $\delta$. In each interval a representative
energy level is selected, with its energy $\varepsilon_{n}$ chosen
as the arithmetic mean of the interval boundaries in the linear sector
(below $D^{*}$) and the geometric mean in the logarithmic sector
(above $D^{*}$). The coupling Hamiltonian is then integrated over
each interval in order to derive the appropriate coupling $v_{\alpha n\lambda}$
of the new lead level with the impurity $\lambda$ level: 
\begin{align}
\varepsilon_{n} & =\begin{cases}
\frac{f\left(\left|n\right|+1+z\right)-f\text{\ensuremath{\left(\left|n\right|+z\right)}}}{\log\left[f\left(\left|n\right|+1+z\right)/f\left(\left|n\right|+z\right)\right]} & \underset{\,}{\left|\varepsilon_{n}\right|>D^{*}}\\
\frac{f\left(\left|n\right|+1+z\right)+f\text{\ensuremath{\left(\left|n\right|+z\right)}}}{2} & \overset{\,}{\left|\varepsilon_{n}\right|\leq D^{*}}
\end{cases},\\
v_{\alpha n\lambda} & =\sqrt{\tfrac{\Gamma_{\alpha\lambda}}{2\pi D}}\left[f\left(\left|n\right|+1+z\right)-f\text{\ensuremath{\left(\left|n\right|+z\right)}}\right].
\end{align}

\section{Exact Solution of the Continuous Noninteracting Case\label{sec:Appendix-Exact-Solution}}

The exact solution for a quadratic continuous system can be calculated
in the Keldysh formalism. For noninteracting leads, all the effects
of the couplings to the leads on the impurity are encoded in the hybridization
function, defined between the $\lambda$th and $\nu$th impurity levels
($\lambda,\nu\in\left\{ 1,\ldots,m\right\} $) for lead $\alpha\in\left\{ L,R\right\} $
as:
\begin{equation}
\Gamma_{\alpha}^{\lambda\nu}\left(\omega\right)=\pi\sum_{n}v_{\alpha n\lambda}^{*}v_{\alpha n\nu}\,\delta\left(\varepsilon_{n}-\omega\right),
\end{equation}
where $v_{\alpha n\lambda}$ are the coupling constants between the
$\lambda$th impurity level and the $n$th energy level of lead $\alpha$
(in the \textit{star} geometry). In the case of continuous leads,
the sum over dense levels $\varepsilon_{n}$ is understood as an integral
over the energies. The total hybridization function is then defined
as a sum on the hybridization functions of all leads: 
\begin{equation}
\Gamma^{\lambda\nu}\left(\omega\right)=\sum_{\alpha}\Gamma_{\alpha}^{\lambda\nu}\left(\omega\right).
\end{equation}
For a quadratic dot Hamiltonian $H$, the retarded and advanced Green's
functions of the dressed impurity are then given by:

\begin{align}
G^{R}\left(\omega\right) & =\left(\omega-H+i\Gamma\left(\omega\right)\right)^{-1},\label{eq:Retarded-Advanced-Greens}\\
G^{A}\left(\omega\right) & =\left(\omega-H-i\Gamma\left(\omega\right)\right)^{-1},\nonumber 
\end{align}
where $H$ and $\Gamma$ are understood here to be $m\times m$ matrices.
The NESS current can be obtained by the Meir-Wingreen formula \citep{Meir92},
which for equal hybridization functions $\Gamma_{L}\left(\omega\right)=\Gamma_{R}\left(\omega\right)=\Gamma\left(\omega\right)/2$,
simplifies to: 
\begin{align}
I & =\frac{i}{2}\int d\omega\left(f_{L}\left(\omega\right)-f_{R}\left(\omega\right)\right)\label{eq:Meir-Wingreen-Current}\\
 & \qquad\qquad\qquad\times\mathrm{tr}\left\{ \Gamma\left(\omega\right)\left(G^{R}\left(\omega\right)-G^{A}\left(\omega\right)\right)\right\} ,\nonumber 
\end{align}
where $f_{\alpha}\left(\omega\right)$ is the lead specific Fermi-Dirac
distribution. The Keldysh Green's function equals:
\begin{equation}
G^{K}\left(\omega\right)=-2i\sum_{\alpha}\left(1-2f_{\alpha}\left(\omega\right)\right)G^{R}\left(\omega\right)\Gamma_{\alpha}\left(\omega\right)G^{A}\left(\omega\right),\label{eq:Keldysh-Greens}
\end{equation}
and the impurity single-particle density matrix can then be obtained
by integrating over it:
\begin{align}
\left\langle d_{\lambda}^{\dagger}d_{\nu}\right\rangle  & =\frac{1}{2}\left(\delta_{\lambda\nu}-\left\langle \left[d_{\nu},d_{\lambda}^{\dagger}\right]\right\rangle \right)\label{eq:Single-Particle-Density-Matrix}\\
 & =\frac{1}{2}\left(\delta_{\lambda\nu}-\frac{i}{2\pi}\int d\omega G_{\nu\lambda}^{K}\left(\omega\right)\right).\nonumber 
\end{align}
Specifying a box hybridization function for the $\lambda$th level
with half bandwidth $D$:
\begin{equation}
\Gamma^{\lambda\lambda}\left(\omega\right)=\Gamma^{\lambda\lambda}\,\Theta\left(D-\left|\omega\right|\right),
\end{equation}
and taking coupling constants $v_{\alpha n\lambda}=v$ which are lead,
$n$ and $\lambda$ independent (so that all the elements of $\Gamma\left(\omega\right)$
are equal), as is indeed the case for both models under investigation
in the continuum limit, Eq.~(\ref{eq:Meir-Wingreen-Current}) simplifies
to:
\begin{equation}
I=\frac{i}{2}\int_{-\frac{V}{2}}^{\frac{V}{2}}d\omega\mathrm{tr}\left\{ \Gamma\,\left(G^{R}\left(\omega\right)-G^{A}\left(\omega\right)\right)\right\} ,
\end{equation}
which can then be evaluated for any desired bias voltage. The Keldysh
Green's function in Eq.~(\ref{eq:Keldysh-Greens}) also simplifies
to: 
\begin{align}
G^{K}\left(\omega\right) & =-2i\Bigl(1-\sum_{\alpha}f_{\alpha}\left(\omega\right)\Bigr)G^{R}\left(\omega\right)\Gamma\left(\omega\right)G^{A}\left(\omega\right)\nonumber \\
 & =\begin{cases}
+2iG^{R}\left(\omega\right)\Gamma\,G^{A}\left(\omega\right) & -D<\omega<-\frac{V}{2}\\
-2iG^{R}\left(\omega\right)\Gamma\,G^{A}\left(\omega\right) & +\frac{V}{2}<\omega<+D\\
\qquad\qquad0 & \qquad\ \mathrm{else}
\end{cases},\label{eq: Simplifed-Keldysh-Greens}
\end{align}
resulting in a simple integral for the single-particle density matrix.

\section{Exact Evolution of the Discrete Noninteracting Case: The Lyapunov
Equation\label{sec:Appendix-Lyapunuv-Equation}}

The single-particle single-time correlation matrix $P_{rs}\left(t\right)\equiv\left\langle c_{r}\left(t\right)c_{s}^{\dagger}\left(t\right)\right\rangle $
encodes all information regarding single-time observables of interest
in this paper, e.g., the impurity current. Furthermore, for quadratic
systems, this matrix encodes all information about the state of the
system, so that finding $P\left(t\right)$ amounts to fully solving
the system. In the case of a quadratic Lindblad equation (both in
the Hamiltonian and the dissipative terms), the exact evolution, as
well as the steady-state solution, can be reduced to a continuous
Lyapunov equation for $P$. The key parts of this reduction are derived
in this appendix, following Ref.~\citep{Schwarz16}. We start from
the most general Lindblad equation for fermionic Lindblad operators
$\left\{ c_{q}\right\} $:
\begin{align}
\frac{\partial\rho}{\partial t}=-i\left[H,\rho\right] & +\sum_{mn}\Lambda_{mn}^{\left(1\right)}\left(2c_{n}\rho c_{m}^{\dagger}-\left\{ c_{m}^{\dagger}c_{n},\rho\right\} \right)\\
 & +\sum_{mn}\Lambda_{mn}^{\left(2\right)}\left(2c_{m}^{\dagger}\rho c_{n}-\left\{ c_{n}c_{m}^{\dagger},\rho\right\} \right),\nonumber 
\end{align}
where $\Lambda^{\left(1,2\right)}$ encode the Lindblad driving rates.
The time dependence of a general single-time observable $\left\langle A\left(t\right)\right\rangle \equiv\mathrm{tr}\left(A\rho\left(t\right)\right)$
is then given by:
\begin{align}
\frac{d\left\langle A\right\rangle }{dt}=-i\left\langle \left[A,H\right]\right\rangle  & +\sum_{mn}\Lambda_{mn}^{\left(1\right)}\left\langle 2c_{m}^{\dagger}Ac_{n}-\left\{ c_{m}^{\dagger}c_{n},A\right\} \right\rangle \label{eq:single-time-observable}\\
 & +\sum_{mn}\Lambda_{mn}^{\left(2\right)}\left\langle 2c_{n}Ac_{m}^{\dagger}-\left\{ c_{n}c_{m}^{\dagger},A\right\} \right\rangle .\nonumber 
\end{align}
Assuming a quadratic Hamiltonian $H=\sum_{mn}H_{mn}c_{m}^{\dagger}c_{n}$,
and substituting $\left\langle A\right\rangle =P_{rs}$ into Eq.~(\ref{eq:single-time-observable}),
results in a differential continuous Lyapunov equation for $P$:
\begin{align}
\frac{dP}{dt} & =AP+PA^{\dagger}+M,\label{eq:diff-Lyapunov}\\
A & \equiv-iH-\Lambda^{\left(1\right)}-\Lambda^{\left(2\right)},\qquad M\equiv2\Lambda^{\left(1\right)}.\nonumber 
\end{align}
The general solution of this equation, for some initial condition
$P_{0}$, is: 
\begin{align}
P\left(t\right) & =e^{At}P_{0}e^{A^{\dagger}t}+\int_{0}^{t}e^{At^{\prime}}Me^{A^{\dagger}t^{\prime}}dt^{\prime}.\label{eq:diff-Lyapunov-sol}
\end{align}
By diagonalizing $A$ (if possible) the integral can be explicitly
calculated, resulting in a closed expression for $P\left(t\right)$.
The steady-state solution is given by $P_{S}$ satisfying $\frac{dP_{S}}{dt}=0$.
It can be obtained by solving the algebraic continuous Lyapunov equation
\begin{equation}
AP_{S}+P_{S}A^{\dagger}+M=0.
\end{equation}
Exploiting the fact that by construction $A,A^{\dagger}$ have only
eigenvalues with a non-positive real part, the steady-state solution
can also be obtained by taking the infinite time limit of Eq.~(\ref{eq:diff-Lyapunov-sol})
\begin{equation}
P_{S}=P\left(t\rightarrow\infty\right)=\int_{0}^{\infty}e^{At^{\prime}}Me^{A^{\dagger}t^{\prime}}dt^{\prime}.
\end{equation}

\begin{figure*}[t]
\begin{centering}
\includegraphics[width=1\textwidth]{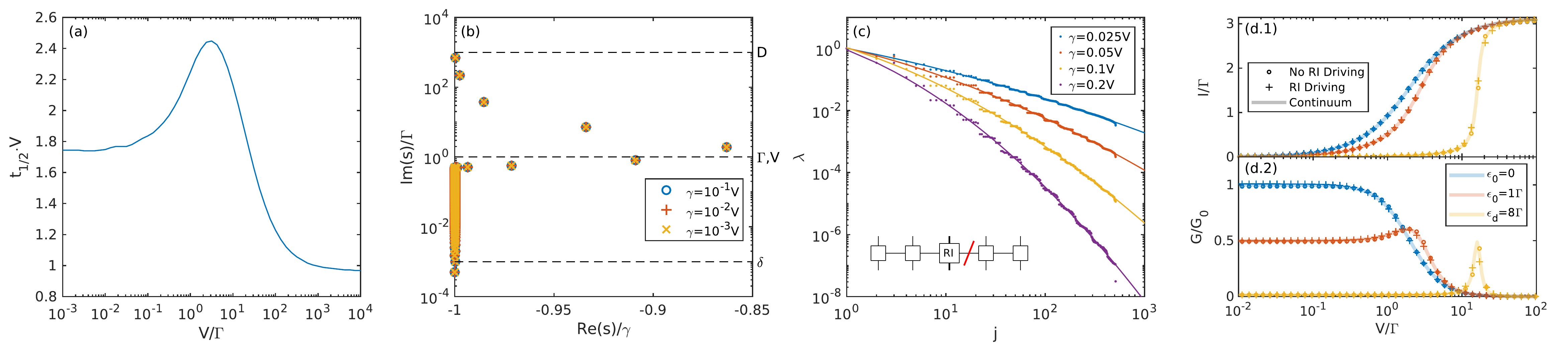}
\par\end{centering}
\caption{(a) The time $t_{1/2}$ (multiplied by $V$) at which the current
reaches half its final value, plotted for the RLM with $\varepsilon_{0}=0$
over a wide range of bias voltages. (b) The distribution in the complex
plane of the eigenvalues of the matrix $A=-iH-\Lambda^{\left(1\right)}-\Lambda^{\left(2\right)}$,
defined in Eq.~(\ref{eq:diff-Lyapunov}), for the RLM with several
choices of driving rates $\gamma$. The imaginary parts, mostly corresponding
to the Hamiltonian, are plotted in units of $\Gamma$, while the real
parts, which are related to the Lindblad driving, are rescaled by
$\gamma$. The closely bunched points near $\mathrm{Re\left(s\right)=-\gamma}$
correspond approximately to the single-particle energies of the Hamiltonian
arising from the linear sector. (c) Example of the long-time limit
steady-state singular value spectrum of the MPDO bond connecting the
RI to one of the linear sector leads (as indicated by the red line
in the cartoon). The spectrum was obtained for the RLM with $V=1,\ \delta=0.025V,\ \chi=512,\ R=32$
and several values of $\gamma$. The singular values were rescaled
such that the largest singular value for each $\gamma$ is 1, and
fitted to a log-Gaussian (solid line). (d) NESS current and differential
conductance of the RLM as a function of bias voltage $V$ and at different
gate voltages $\varepsilon_{0}$, as calculated with (pluses) and
without (circles) Lindblad driving of the sites enclosed in the RI,
and compared with the continuum limit (shaded). The current with and
without the driving at the RI is calculated exactly for $\gamma/\delta=2,1$
(with $\delta=0.05V$) and $\Lambda=8,6$, and is then linearly extrapolated
to $\gamma\rightarrow0,\ \Lambda\rightarrow1$.\label{fig:Appendices}}
\end{figure*}

\section{Evolution Time Scales\label{sec:Appendix-Evolution-Time-Scales}}

In this section we analyze the time scales of the current evolution
for the RLM with $\varepsilon_{0}=0$, after discretization in the
linear-logarithmic scheme, and with energy independent Lindblad driving
$\gamma$, as discussed in Sec.~\ref{subsec:Lindblad-Equation}.
The time scale of the initial rise in the current can be characterized
by $t_{1/2}$, the time at which the current first reaches half of
its final value. This time scale appears to be inversely proportional
to the bias voltage $V$, as can be seen in Fig.~\ref{fig:Appendices}(a)
where $t_{1/2}\cdot V$ is of order unity over the full range of explored
bias.

The time scale of the decay towards the steady-state can be extracted
for a quadratic model from the matrix $A$, defined in Eq.~(\ref{eq:diff-Lyapunov}).
The (negative) real parts of the eigenvalues of this matrix dictate
the decay rate of each mode. The ones with the smallest magnitude
set a bound on the total decay rate of the system. For sufficiently
small Lindblad driving, the imaginary part of the eigenvalues depends
mainly on the Hamiltonian, while the real part will depend on the
driving rates. Thus for the RLM in the discussed discretization scheme,
the real part of the eigenvalues naturally scales with $\gamma$,
as can be seen in Fig.~\ref{fig:Appendices}(b) for several choices
of $\gamma$, and the decay rate is proportional to $\gamma$, with
the proportionality constant of order 1.

\section{MPDO Singular Value Spectrum\label{sec:MPDO-Singular-Values}}

In this section we discuss the dependence of the long-time limit steady-state
singular value spectrum of the MPDO on the Lindblad driving rate $\gamma$.
As an example we plot in Fig.~\ref{fig:Appendices}(c) the singular
value spectrum, taken at the bond connecting the RI to one of the
linear sector leads (as indicated in the cartoon), for the RLM with
parameters as given in the caption. First note that while the normalization
of the wavefunction constraints the squared singular values of an
MPS to sum up to 1, the density operator normalization condition does
not impose any constraint on the MPDO singular values. Thus the global
prefactor is arbitrary, and for clarity the singular values are rescaled
such that the largest singular value for each $\gamma$ is 1. As can
be seen in the figure, the singular values decay at a faster than
power-law rate, implying an efficient representation of the steady-state
as an MPDO with finite bond dimension $\chi$. Moreover, we observe
that the decay rate grows monotonically with increasing $\gamma$,
implying that larger $\gamma$ requires a smaller bond dimension in
order to efficiently represent the state of the system.

A full characterization of the exact functional dependence of the
singular values $\lambda_{j}$ on the index $j$ and the system parameters
requires a more detailed analysis than carried out in this work. We
do note however, that we can fit it to a log-Gaussian behavior $\lambda_{j}\propto e^{-\left(a\log j+b\right)^{2}}$,
with $a$ and $b$ the fitting parameters. We suspect that this specific
behavior for an MPDO steady-state is not coincidental, since a similar
behavior has been argued to occur for an MPS ground-state \citep{Calabrese08}.
We further observe that the fitting parameter $a$, which dictates
the decay rate, is monotonic in $\gamma$ and goes to zero in the
$\gamma\rightarrow0$ limit. Thus in this limit the required bond
dimension $\chi$ diverges. This is to be expected, as the steady-state
corresponds to evolution to infinite time without dissipation, and
we get the well known exponential growth in entanglement entropy for
unitary evolution.

\section{Driving RI Sites\label{sec:Appendix-RI-Driving}}

Fig.~\ref{fig:Appendices}(d) demonstrates that Lindblad driving
of the RI itself has a negligible effect on the resulting current
and differential conductance, with respect to an exact solution (which
is attainable for the RLM). As argued in Sec.~\ref{sec:Error-Analysis},
this is because the RI represents energy levels far from the voltage
or temperature bias scales. These levels are not expected to be affected
by the nonequilibrium conditions and thus only set the (renormalized)
stage for the low-energy dynamics. Moreover, the exact solution of
the modified Lindblad equation (without driving the RI) is still a
valid approximation for the continuous system in the limits $\Lambda\rightarrow1,\ \gamma=\delta\rightarrow0$.
This justifies turning off the driving for the interacting case, thus
suppressing numerical artifacts arising due to the interplay between
NRG and the dissipative dynamics.

\bibliographystyle{apsrev4-1}
\bibliography{bib}

\begin{thebibliography}{60}%
\makeatletter
\providecommand \@ifxundefined [1]{%
 \@ifx{#1\undefined}
}%
\providecommand \@ifnum [1]{%
 \ifnum #1\expandafter \@firstoftwo
 \else \expandafter \@secondoftwo
 \fi
}%
\providecommand \@ifx [1]{%
 \ifx #1\expandafter \@firstoftwo
 \else \expandafter \@secondoftwo
 \fi
}%
\providecommand \natexlab [1]{#1}%
\providecommand \enquote  [1]{``#1''}%
\providecommand \bibnamefont  [1]{#1}%
\providecommand \bibfnamefont [1]{#1}%
\providecommand \citenamefont [1]{#1}%
\providecommand \href@noop [0]{\@secondoftwo}%
\providecommand \href [0]{\begingroup \@sanitize@url \@href}%
\providecommand \@href[1]{\@@startlink{#1}\@@href}%
\providecommand \@@href[1]{\endgroup#1\@@endlink}%
\providecommand \@sanitize@url [0]{\catcode `\\12\catcode `\$12\catcode
  `\&12\catcode `\#12\catcode `\^12\catcode `\_12\catcode `\%12\relax}%
\providecommand \@@startlink[1]{}%
\providecommand \@@endlink[0]{}%
\providecommand \url  [0]{\begingroup\@sanitize@url \@url }%
\providecommand \@url [1]{\endgroup\@href {#1}{\urlprefix }}%
\providecommand \urlprefix  [0]{URL }%
\providecommand \Eprint [0]{\href }%
\providecommand \doibase [0]{http://dx.doi.org/}%
\providecommand \selectlanguage [0]{\@gobble}%
\providecommand \bibinfo  [0]{\@secondoftwo}%
\providecommand \bibfield  [0]{\@secondoftwo}%
\providecommand \translation [1]{[#1]}%
\providecommand \BibitemOpen [0]{}%
\providecommand \bibitemStop [0]{}%
\providecommand \bibitemNoStop [0]{.\EOS\space}%
\providecommand \EOS [0]{\spacefactor3000\relax}%
\providecommand \BibitemShut  [1]{\csname bibitem#1\endcsname}%
\let\auto@bib@innerbib\@empty
%</preamble>
\bibitem [{\citenamefont {Kondo}(1964)}]{Kondo64}%
  \BibitemOpen
  \bibfield  {author} {\bibinfo {author} {\bibfnamefont {J.}~\bibnamefont
  {Kondo}},\ }\href {\doibase 10.1143/PTP.32.37} {\bibfield  {journal}
  {\bibinfo  {journal} {Progress of Theoretical Physics}\ }\textbf {\bibinfo
  {volume} {32}},\ \bibinfo {pages} {37} (\bibinfo {year} {1964})}\BibitemShut
  {NoStop}%
\bibitem [{\citenamefont {Hewson}(1993)}]{hewson_1993}%
  \BibitemOpen
  \bibfield  {author} {\bibinfo {author} {\bibfnamefont {A.~C.}\ \bibnamefont
  {Hewson}},\ }\href {\doibase 10.1017/CBO9780511470752} {\emph {\bibinfo
  {title} {The Kondo Problem to Heavy Fermions}}},\ Cambridge Studies in
  Magnetism\ (\bibinfo  {publisher} {Cambridge University Press},\ \bibinfo
  {year} {1993})\BibitemShut {NoStop}%
\bibitem [{\citenamefont {Wilson}(1975)}]{Wilson75}%
  \BibitemOpen
  \bibfield  {author} {\bibinfo {author} {\bibfnamefont {K.~G.}\ \bibnamefont
  {Wilson}},\ }\href {\doibase 10.1103/RevModPhys.47.773} {\bibfield  {journal}
  {\bibinfo  {journal} {Rev. Mod. Phys.}\ }\textbf {\bibinfo {volume} {47}},\
  \bibinfo {pages} {773} (\bibinfo {year} {1975})}\BibitemShut {NoStop}%
\bibitem [{\citenamefont {Bulla}\ \emph {et~al.}(2008)\citenamefont {Bulla},
  \citenamefont {Costi},\ and\ \citenamefont {Pruschke}}]{Bulla08}%
  \BibitemOpen
  \bibfield  {author} {\bibinfo {author} {\bibfnamefont {R.}~\bibnamefont
  {Bulla}}, \bibinfo {author} {\bibfnamefont {T.~A.}\ \bibnamefont {Costi}}, \
  and\ \bibinfo {author} {\bibfnamefont {T.}~\bibnamefont {Pruschke}},\ }\href
  {\doibase 10.1103/RevModPhys.80.395} {\bibfield  {journal} {\bibinfo
  {journal} {Rev. Mod. Phys.}\ }\textbf {\bibinfo {volume} {80}},\ \bibinfo
  {pages} {395} (\bibinfo {year} {2008})}\BibitemShut {NoStop}%
\bibitem [{\citenamefont {Goldhaber-Gordon}\ \emph {et~al.}(1998)\citenamefont
  {Goldhaber-Gordon}, \citenamefont {Shtrikman}, \citenamefont {Mahalu},
  \citenamefont {Abusch-Magder}, \citenamefont {Meirav},\ and\ \citenamefont
  {Kastner}}]{Goldhaber-Gordon1998}%
  \BibitemOpen
  \bibfield  {author} {\bibinfo {author} {\bibfnamefont {D.}~\bibnamefont
  {Goldhaber-Gordon}}, \bibinfo {author} {\bibfnamefont {H.}~\bibnamefont
  {Shtrikman}}, \bibinfo {author} {\bibfnamefont {D.}~\bibnamefont {Mahalu}},
  \bibinfo {author} {\bibfnamefont {D.}~\bibnamefont {Abusch-Magder}}, \bibinfo
  {author} {\bibfnamefont {U.}~\bibnamefont {Meirav}}, \ and\ \bibinfo {author}
  {\bibfnamefont {M.~A.}\ \bibnamefont {Kastner}},\ }\href {\doibase
  10.1038/34373} {\bibfield  {journal} {\bibinfo  {journal} {Nature}\ }\textbf
  {\bibinfo {volume} {391}},\ \bibinfo {pages} {156} (\bibinfo {year}
  {1998})}\BibitemShut {NoStop}%
\bibitem [{\citenamefont {Cronenwett}\ \emph {et~al.}(1998)\citenamefont
  {Cronenwett}, \citenamefont {Oosterkamp},\ and\ \citenamefont
  {Kouwenhoven}}]{Cronenwett540}%
  \BibitemOpen
  \bibfield  {author} {\bibinfo {author} {\bibfnamefont {S.~M.}\ \bibnamefont
  {Cronenwett}}, \bibinfo {author} {\bibfnamefont {T.~H.}\ \bibnamefont
  {Oosterkamp}}, \ and\ \bibinfo {author} {\bibfnamefont {L.~P.}\ \bibnamefont
  {Kouwenhoven}},\ }\href {\doibase 10.1126/science.281.5376.540} {\bibfield
  {journal} {\bibinfo  {journal} {Science}\ }\textbf {\bibinfo {volume}
  {281}},\ \bibinfo {pages} {540} (\bibinfo {year} {1998})}\BibitemShut
  {NoStop}%
\bibitem [{\citenamefont {Nyg{\aa}rd}\ \emph {et~al.}(2000)\citenamefont
  {Nyg{\aa}rd}, \citenamefont {Cobden},\ and\ \citenamefont
  {Lindelof}}]{Nygard2000}%
  \BibitemOpen
  \bibfield  {author} {\bibinfo {author} {\bibfnamefont {J.}~\bibnamefont
  {Nyg{\aa}rd}}, \bibinfo {author} {\bibfnamefont {D.~H.}\ \bibnamefont
  {Cobden}}, \ and\ \bibinfo {author} {\bibfnamefont {P.~E.}\ \bibnamefont
  {Lindelof}},\ }\href {\doibase 10.1038/35042545} {\bibfield  {journal}
  {\bibinfo  {journal} {Nature}\ }\textbf {\bibinfo {volume} {408}},\ \bibinfo
  {pages} {342} (\bibinfo {year} {2000})}\BibitemShut {NoStop}%
\bibitem [{\citenamefont {Buitelaar}\ \emph {et~al.}(2002)\citenamefont
  {Buitelaar}, \citenamefont {Bachtold}, \citenamefont {Nussbaumer},
  \citenamefont {Iqbal},\ and\ \citenamefont {Sch\"onenberger}}]{Buitelaar02}%
  \BibitemOpen
  \bibfield  {author} {\bibinfo {author} {\bibfnamefont {M.~R.}\ \bibnamefont
  {Buitelaar}}, \bibinfo {author} {\bibfnamefont {A.}~\bibnamefont {Bachtold}},
  \bibinfo {author} {\bibfnamefont {T.}~\bibnamefont {Nussbaumer}}, \bibinfo
  {author} {\bibfnamefont {M.}~\bibnamefont {Iqbal}}, \ and\ \bibinfo {author}
  {\bibfnamefont {C.}~\bibnamefont {Sch\"onenberger}},\ }\href {\doibase
  10.1103/PhysRevLett.88.156801} {\bibfield  {journal} {\bibinfo  {journal}
  {Phys. Rev. Lett.}\ }\textbf {\bibinfo {volume} {88}},\ \bibinfo {pages}
  {156801} (\bibinfo {year} {2002})}\BibitemShut {NoStop}%
\bibitem [{\citenamefont {Park}\ \emph {et~al.}(2002)\citenamefont {Park},
  \citenamefont {Pasupathy}, \citenamefont {Goldsmith}, \citenamefont {Chang},
  \citenamefont {Yaish}, \citenamefont {Petta}, \citenamefont {Rinkoski},
  \citenamefont {Sethna}, \citenamefont {Abru{\~n}a}, \citenamefont {McEuen},\
  and\ \citenamefont {Ralph}}]{Park2002}%
  \BibitemOpen
  \bibfield  {author} {\bibinfo {author} {\bibfnamefont {J.}~\bibnamefont
  {Park}}, \bibinfo {author} {\bibfnamefont {A.~N.}\ \bibnamefont {Pasupathy}},
  \bibinfo {author} {\bibfnamefont {J.~I.}\ \bibnamefont {Goldsmith}}, \bibinfo
  {author} {\bibfnamefont {C.}~\bibnamefont {Chang}}, \bibinfo {author}
  {\bibfnamefont {Y.}~\bibnamefont {Yaish}}, \bibinfo {author} {\bibfnamefont
  {J.~R.}\ \bibnamefont {Petta}}, \bibinfo {author} {\bibfnamefont
  {M.}~\bibnamefont {Rinkoski}}, \bibinfo {author} {\bibfnamefont {J.~P.}\
  \bibnamefont {Sethna}}, \bibinfo {author} {\bibfnamefont {H.~D.}\
  \bibnamefont {Abru{\~n}a}}, \bibinfo {author} {\bibfnamefont {P.~L.}\
  \bibnamefont {McEuen}}, \ and\ \bibinfo {author} {\bibfnamefont {D.~C.}\
  \bibnamefont {Ralph}},\ }\href {\doibase 10.1038/nature00791} {\bibfield
  {journal} {\bibinfo  {journal} {Nature}\ }\textbf {\bibinfo {volume} {417}},\
  \bibinfo {pages} {722} (\bibinfo {year} {2002})}\BibitemShut {NoStop}%
\bibitem [{\citenamefont {Liang}\ \emph {et~al.}(2002)\citenamefont {Liang},
  \citenamefont {Shores}, \citenamefont {Bockrath}, \citenamefont {Long},\ and\
  \citenamefont {Park}}]{Liang2002}%
  \BibitemOpen
  \bibfield  {author} {\bibinfo {author} {\bibfnamefont {W.}~\bibnamefont
  {Liang}}, \bibinfo {author} {\bibfnamefont {M.~P.}\ \bibnamefont {Shores}},
  \bibinfo {author} {\bibfnamefont {M.}~\bibnamefont {Bockrath}}, \bibinfo
  {author} {\bibfnamefont {J.~R.}\ \bibnamefont {Long}}, \ and\ \bibinfo
  {author} {\bibfnamefont {H.}~\bibnamefont {Park}},\ }\href {\doibase
  10.1038/nature00790} {\bibfield  {journal} {\bibinfo  {journal} {Nature}\
  }\textbf {\bibinfo {volume} {417}},\ \bibinfo {pages} {725} (\bibinfo {year}
  {2002})}\BibitemShut {NoStop}%
\bibitem [{\citenamefont {Beenakker}(1991)}]{Beenakker91}%
  \BibitemOpen
  \bibfield  {author} {\bibinfo {author} {\bibfnamefont {C.~W.~J.}\
  \bibnamefont {Beenakker}},\ }\href {\doibase 10.1103/PhysRevB.44.1646}
  {\bibfield  {journal} {\bibinfo  {journal} {Phys. Rev. B}\ }\textbf {\bibinfo
  {volume} {44}},\ \bibinfo {pages} {1646} (\bibinfo {year}
  {1991})}\BibitemShut {NoStop}%
\bibitem [{\citenamefont {Rosch}\ \emph {et~al.}(2003)\citenamefont {Rosch},
  \citenamefont {Paaske}, \citenamefont {Kroha},\ and\ \citenamefont
  {W\"olfle}}]{Rosch03}%
  \BibitemOpen
  \bibfield  {author} {\bibinfo {author} {\bibfnamefont {A.}~\bibnamefont
  {Rosch}}, \bibinfo {author} {\bibfnamefont {J.}~\bibnamefont {Paaske}},
  \bibinfo {author} {\bibfnamefont {J.}~\bibnamefont {Kroha}}, \ and\ \bibinfo
  {author} {\bibfnamefont {P.}~\bibnamefont {W\"olfle}},\ }\href {\doibase
  10.1103/PhysRevLett.90.076804} {\bibfield  {journal} {\bibinfo  {journal}
  {Phys. Rev. Lett.}\ }\textbf {\bibinfo {volume} {90}},\ \bibinfo {pages}
  {076804} (\bibinfo {year} {2003})}\BibitemShut {NoStop}%
\bibitem [{\citenamefont {Kehrein}(2005)}]{Kehrein05}%
  \BibitemOpen
  \bibfield  {author} {\bibinfo {author} {\bibfnamefont {S.}~\bibnamefont
  {Kehrein}},\ }\href {\doibase 10.1103/PhysRevLett.95.056602} {\bibfield
  {journal} {\bibinfo  {journal} {Phys. Rev. Lett.}\ }\textbf {\bibinfo
  {volume} {95}},\ \bibinfo {pages} {056602} (\bibinfo {year}
  {2005})}\BibitemShut {NoStop}%
\bibitem [{\citenamefont {Pletyukhov}\ and\ \citenamefont
  {Schoeller}(2012)}]{Pletyukhov12}%
  \BibitemOpen
  \bibfield  {author} {\bibinfo {author} {\bibfnamefont {M.}~\bibnamefont
  {Pletyukhov}}\ and\ \bibinfo {author} {\bibfnamefont {H.}~\bibnamefont
  {Schoeller}},\ }\href {\doibase 10.1103/PhysRevLett.108.260601} {\bibfield
  {journal} {\bibinfo  {journal} {Phys. Rev. Lett.}\ }\textbf {\bibinfo
  {volume} {108}},\ \bibinfo {pages} {260601} (\bibinfo {year}
  {2012})}\BibitemShut {NoStop}%
\bibitem [{\citenamefont {Smirnov}\ and\ \citenamefont
  {Grifoni}(2013)}]{Smirnov13}%
  \BibitemOpen
  \bibfield  {author} {\bibinfo {author} {\bibfnamefont {S.}~\bibnamefont
  {Smirnov}}\ and\ \bibinfo {author} {\bibfnamefont {M.}~\bibnamefont
  {Grifoni}},\ }\href {\doibase 10.1103/PhysRevB.87.121302} {\bibfield
  {journal} {\bibinfo  {journal} {Phys. Rev. B}\ }\textbf {\bibinfo {volume}
  {87}},\ \bibinfo {pages} {121302} (\bibinfo {year} {2013})}\BibitemShut
  {NoStop}%
\bibitem [{\citenamefont {Mehta}\ and\ \citenamefont {Andrei}(2006)}]{Mehta06}%
  \BibitemOpen
  \bibfield  {author} {\bibinfo {author} {\bibfnamefont {P.}~\bibnamefont
  {Mehta}}\ and\ \bibinfo {author} {\bibfnamefont {N.}~\bibnamefont {Andrei}},\
  }\href {\doibase 10.1103/PhysRevLett.96.216802} {\bibfield  {journal}
  {\bibinfo  {journal} {Phys. Rev. Lett.}\ }\textbf {\bibinfo {volume} {96}},\
  \bibinfo {pages} {216802} (\bibinfo {year} {2006})}\BibitemShut {NoStop}%
\bibitem [{\citenamefont {Culver}\ and\ \citenamefont
  {Andrei}(2019)}]{culver2019nonperturbative}%
  \BibitemOpen
  \bibfield  {author} {\bibinfo {author} {\bibfnamefont {A.~B.}\ \bibnamefont
  {Culver}}\ and\ \bibinfo {author} {\bibfnamefont {N.}~\bibnamefont
  {Andrei}},\ }\href@noop {} {\  (\bibinfo {year} {2019})},\ \Eprint
  {http://arxiv.org/abs/1912.02956} {arXiv:1912.02956 [cond-mat.str-el]}
  \BibitemShut {NoStop}%
\bibitem [{\citenamefont {Boulat}\ \emph {et~al.}(2008)\citenamefont {Boulat},
  \citenamefont {Saleur},\ and\ \citenamefont {Schmitteckert}}]{Boulat08}%
  \BibitemOpen
  \bibfield  {author} {\bibinfo {author} {\bibfnamefont {E.}~\bibnamefont
  {Boulat}}, \bibinfo {author} {\bibfnamefont {H.}~\bibnamefont {Saleur}}, \
  and\ \bibinfo {author} {\bibfnamefont {P.}~\bibnamefont {Schmitteckert}},\
  }\href {\doibase 10.1103/PhysRevLett.101.140601} {\bibfield  {journal}
  {\bibinfo  {journal} {Phys. Rev. Lett.}\ }\textbf {\bibinfo {volume} {101}},\
  \bibinfo {pages} {140601} (\bibinfo {year} {2008})}\BibitemShut {NoStop}%
\bibitem [{\citenamefont {Dias~da Silva}\ \emph {et~al.}(2008)\citenamefont
  {Dias~da Silva}, \citenamefont {Heidrich-Meisner}, \citenamefont {Feiguin},
  \citenamefont {B\"usser}, \citenamefont {Martins}, \citenamefont {Anda},\
  and\ \citenamefont {Dagotto}}]{daSilva08}%
  \BibitemOpen
  \bibfield  {author} {\bibinfo {author} {\bibfnamefont {L.~G. G.~V.}\
  \bibnamefont {Dias~da Silva}}, \bibinfo {author} {\bibfnamefont
  {F.}~\bibnamefont {Heidrich-Meisner}}, \bibinfo {author} {\bibfnamefont
  {A.~E.}\ \bibnamefont {Feiguin}}, \bibinfo {author} {\bibfnamefont {C.~A.}\
  \bibnamefont {B\"usser}}, \bibinfo {author} {\bibfnamefont {G.~B.}\
  \bibnamefont {Martins}}, \bibinfo {author} {\bibfnamefont {E.~V.}\
  \bibnamefont {Anda}}, \ and\ \bibinfo {author} {\bibfnamefont
  {E.}~\bibnamefont {Dagotto}},\ }\href {\doibase 10.1103/PhysRevB.78.195317}
  {\bibfield  {journal} {\bibinfo  {journal} {Phys. Rev. B}\ }\textbf {\bibinfo
  {volume} {78}},\ \bibinfo {pages} {195317} (\bibinfo {year}
  {2008})}\BibitemShut {NoStop}%
\bibitem [{\citenamefont {Eckel}\ \emph {et~al.}(2010)\citenamefont {Eckel},
  \citenamefont {Heidrich-Meisner}, \citenamefont {Jakobs}, \citenamefont
  {Thorwart}, \citenamefont {Pletyukhov},\ and\ \citenamefont
  {Egger}}]{Eckel_2010}%
  \BibitemOpen
  \bibfield  {author} {\bibinfo {author} {\bibfnamefont {J.}~\bibnamefont
  {Eckel}}, \bibinfo {author} {\bibfnamefont {F.}~\bibnamefont
  {Heidrich-Meisner}}, \bibinfo {author} {\bibfnamefont {S.~G.}\ \bibnamefont
  {Jakobs}}, \bibinfo {author} {\bibfnamefont {M.}~\bibnamefont {Thorwart}},
  \bibinfo {author} {\bibfnamefont {M.}~\bibnamefont {Pletyukhov}}, \ and\
  \bibinfo {author} {\bibfnamefont {R.}~\bibnamefont {Egger}},\ }\href
  {\doibase 10.1088/1367-2630/12/4/043042} {\bibfield  {journal} {\bibinfo
  {journal} {New Journal of Physics}\ }\textbf {\bibinfo {volume} {12}},\
  \bibinfo {pages} {043042} (\bibinfo {year} {2010})}\BibitemShut {NoStop}%
\bibitem [{\citenamefont {Dorda}\ \emph {et~al.}(2015)\citenamefont {Dorda},
  \citenamefont {Ganahl}, \citenamefont {Evertz}, \citenamefont {von~der
  Linden},\ and\ \citenamefont {Arrigoni}}]{Dorda15}%
  \BibitemOpen
  \bibfield  {author} {\bibinfo {author} {\bibfnamefont {A.}~\bibnamefont
  {Dorda}}, \bibinfo {author} {\bibfnamefont {M.}~\bibnamefont {Ganahl}},
  \bibinfo {author} {\bibfnamefont {H.~G.}\ \bibnamefont {Evertz}}, \bibinfo
  {author} {\bibfnamefont {W.}~\bibnamefont {von~der Linden}}, \ and\ \bibinfo
  {author} {\bibfnamefont {E.}~\bibnamefont {Arrigoni}},\ }\href {\doibase
  10.1103/PhysRevB.92.125145} {\bibfield  {journal} {\bibinfo  {journal} {Phys.
  Rev. B}\ }\textbf {\bibinfo {volume} {92}},\ \bibinfo {pages} {125145}
  (\bibinfo {year} {2015})}\BibitemShut {NoStop}%
\bibitem [{\citenamefont {Brenes}\ \emph {et~al.}(2019)\citenamefont {Brenes},
  \citenamefont {Mendoza-Arenas}, \citenamefont {Purkayastha}, \citenamefont
  {Mitchison}, \citenamefont {Clark},\ and\ \citenamefont
  {Goold}}]{brenes2019tensornetwork}%
  \BibitemOpen
  \bibfield  {author} {\bibinfo {author} {\bibfnamefont {M.}~\bibnamefont
  {Brenes}}, \bibinfo {author} {\bibfnamefont {J.~J.}\ \bibnamefont
  {Mendoza-Arenas}}, \bibinfo {author} {\bibfnamefont {A.}~\bibnamefont
  {Purkayastha}}, \bibinfo {author} {\bibfnamefont {M.~T.}\ \bibnamefont
  {Mitchison}}, \bibinfo {author} {\bibfnamefont {S.~R.}\ \bibnamefont
  {Clark}}, \ and\ \bibinfo {author} {\bibfnamefont {J.}~\bibnamefont
  {Goold}},\ }\href@noop {} {\  (\bibinfo {year} {2019})},\ \Eprint
  {http://arxiv.org/abs/1912.02053} {arXiv:1912.02053 [cond-mat.str-el]}
  \BibitemShut {NoStop}%
\bibitem [{\citenamefont {Hod}\ \emph {et~al.}(2016)\citenamefont {Hod},
  \citenamefont {Rodr\'iguez-Rosario}, \citenamefont {Zelovich},\ and\
  \citenamefont {Frauenheim}}]{Hod16}%
  \BibitemOpen
  \bibfield  {author} {\bibinfo {author} {\bibfnamefont {O.}~\bibnamefont
  {Hod}}, \bibinfo {author} {\bibfnamefont {C.~A.}\ \bibnamefont
  {Rodr\'iguez-Rosario}}, \bibinfo {author} {\bibfnamefont {T.}~\bibnamefont
  {Zelovich}}, \ and\ \bibinfo {author} {\bibfnamefont {T.}~\bibnamefont
  {Frauenheim}},\ }\href {\doibase 10.1021/acs.jpca.5b12212} {\bibfield
  {journal} {\bibinfo  {journal} {The Journal of Physical Chemistry A}\
  }\textbf {\bibinfo {volume} {120}},\ \bibinfo {pages} {3278} (\bibinfo {year}
  {2016})}\BibitemShut {NoStop}%
\bibitem [{\citenamefont {Tanimura}\ and\ \citenamefont
  {Kubo}(1989)}]{Tanimura89}%
  \BibitemOpen
  \bibfield  {author} {\bibinfo {author} {\bibfnamefont {Y.}~\bibnamefont
  {Tanimura}}\ and\ \bibinfo {author} {\bibfnamefont {R.}~\bibnamefont
  {Kubo}},\ }\href {\doibase 10.1143/JPSJ.58.101} {\bibfield  {journal}
  {\bibinfo  {journal} {Journal of the Physical Society of Japan}\ }\textbf
  {\bibinfo {volume} {58}},\ \bibinfo {pages} {101} (\bibinfo {year}
  {1989})}\BibitemShut {NoStop}%
\bibitem [{\citenamefont {an~Yan}\ \emph {et~al.}(2004)\citenamefont {an~Yan},
  \citenamefont {Yang}, \citenamefont {Liu},\ and\ \citenamefont
  {Shao}}]{YAN2004216}%
  \BibitemOpen
  \bibfield  {author} {\bibinfo {author} {\bibfnamefont {Y.}~\bibnamefont
  {an~Yan}}, \bibinfo {author} {\bibfnamefont {F.}~\bibnamefont {Yang}},
  \bibinfo {author} {\bibfnamefont {Y.}~\bibnamefont {Liu}}, \ and\ \bibinfo
  {author} {\bibfnamefont {J.}~\bibnamefont {Shao}},\ }\href {\doibase
  https://doi.org/10.1016/j.cplett.2004.07.036} {\bibfield  {journal} {\bibinfo
   {journal} {Chemical Physics Letters}\ }\textbf {\bibinfo {volume} {395}},\
  \bibinfo {pages} {216 } (\bibinfo {year} {2004})}\BibitemShut {NoStop}%
\bibitem [{\citenamefont {Wang}\ and\ \citenamefont {Thoss}(2008)}]{Wang_2008}%
  \BibitemOpen
  \bibfield  {author} {\bibinfo {author} {\bibfnamefont {H.}~\bibnamefont
  {Wang}}\ and\ \bibinfo {author} {\bibfnamefont {M.}~\bibnamefont {Thoss}},\
  }\href {\doibase 10.1088/1367-2630/10/11/115005} {\bibfield  {journal}
  {\bibinfo  {journal} {New Journal of Physics}\ }\textbf {\bibinfo {volume}
  {10}},\ \bibinfo {pages} {115005} (\bibinfo {year} {2008})}\BibitemShut
  {NoStop}%
\bibitem [{\citenamefont {Cohen}\ \emph {et~al.}(2014)\citenamefont {Cohen},
  \citenamefont {Gull}, \citenamefont {Reichman},\ and\ \citenamefont
  {Millis}}]{Cohen14}%
  \BibitemOpen
  \bibfield  {author} {\bibinfo {author} {\bibfnamefont {G.}~\bibnamefont
  {Cohen}}, \bibinfo {author} {\bibfnamefont {E.}~\bibnamefont {Gull}},
  \bibinfo {author} {\bibfnamefont {D.~R.}\ \bibnamefont {Reichman}}, \ and\
  \bibinfo {author} {\bibfnamefont {A.~J.}\ \bibnamefont {Millis}},\ }\href
  {\doibase 10.1103/PhysRevLett.112.146802} {\bibfield  {journal} {\bibinfo
  {journal} {Phys. Rev. Lett.}\ }\textbf {\bibinfo {volume} {112}},\ \bibinfo
  {pages} {146802} (\bibinfo {year} {2014})}\BibitemShut {NoStop}%
\bibitem [{\citenamefont {Anders}(2008)}]{Anders08}%
  \BibitemOpen
  \bibfield  {author} {\bibinfo {author} {\bibfnamefont {F.~B.}\ \bibnamefont
  {Anders}},\ }\href {\doibase 10.1103/PhysRevLett.101.066804} {\bibfield
  {journal} {\bibinfo  {journal} {Phys. Rev. Lett.}\ }\textbf {\bibinfo
  {volume} {101}},\ \bibinfo {pages} {066804} (\bibinfo {year}
  {2008})}\BibitemShut {NoStop}%
\bibitem [{\citenamefont {Schwarz}\ \emph {et~al.}(2018)\citenamefont
  {Schwarz}, \citenamefont {Weymann}, \citenamefont {von Delft},\ and\
  \citenamefont {Weichselbaum}}]{Schwarz18}%
  \BibitemOpen
  \bibfield  {author} {\bibinfo {author} {\bibfnamefont {F.}~\bibnamefont
  {Schwarz}}, \bibinfo {author} {\bibfnamefont {I.}~\bibnamefont {Weymann}},
  \bibinfo {author} {\bibfnamefont {J.}~\bibnamefont {von Delft}}, \ and\
  \bibinfo {author} {\bibfnamefont {A.}~\bibnamefont {Weichselbaum}},\ }\href
  {\doibase 10.1103/PhysRevLett.121.137702} {\bibfield  {journal} {\bibinfo
  {journal} {Phys. Rev. Lett.}\ }\textbf {\bibinfo {volume} {121}},\ \bibinfo
  {pages} {137702} (\bibinfo {year} {2018})}\BibitemShut {NoStop}%
\bibitem [{\citenamefont {Breuer}\ and\ \citenamefont
  {Petruccione}(2007)}]{Breuer_2007}%
  \BibitemOpen
  \bibfield  {author} {\bibinfo {author} {\bibfnamefont {H.}~\bibnamefont
  {Breuer}}\ and\ \bibinfo {author} {\bibfnamefont {F.}~\bibnamefont
  {Petruccione}},\ }\href {\doibase 10.1093/acprof:oso/9780199213900.001.0001}
  {\emph {\bibinfo {title} {The Theory of Open Quantum Systems}}}\ (\bibinfo
  {publisher} {Oxford University Press},\ \bibinfo {year} {2007})\BibitemShut
  {NoStop}%
\bibitem [{\citenamefont {Schwarz}\ \emph {et~al.}(2016)\citenamefont
  {Schwarz}, \citenamefont {Goldstein}, \citenamefont {Dorda}, \citenamefont
  {Arrigoni}, \citenamefont {Weichselbaum},\ and\ \citenamefont {von
  Delft}}]{Schwarz16}%
  \BibitemOpen
  \bibfield  {author} {\bibinfo {author} {\bibfnamefont {F.}~\bibnamefont
  {Schwarz}}, \bibinfo {author} {\bibfnamefont {M.}~\bibnamefont {Goldstein}},
  \bibinfo {author} {\bibfnamefont {A.}~\bibnamefont {Dorda}}, \bibinfo
  {author} {\bibfnamefont {E.}~\bibnamefont {Arrigoni}}, \bibinfo {author}
  {\bibfnamefont {A.}~\bibnamefont {Weichselbaum}}, \ and\ \bibinfo {author}
  {\bibfnamefont {J.}~\bibnamefont {von Delft}},\ }\href {\doibase
  10.1103/PhysRevB.94.155142} {\bibfield  {journal} {\bibinfo  {journal} {Phys.
  Rev. B}\ }\textbf {\bibinfo {volume} {94}},\ \bibinfo {pages} {155142}
  (\bibinfo {year} {2016})}\BibitemShut {NoStop}%
\bibitem [{\citenamefont {White}(1992)}]{White92}%
  \BibitemOpen
  \bibfield  {author} {\bibinfo {author} {\bibfnamefont {S.~R.}\ \bibnamefont
  {White}},\ }\href {\doibase 10.1103/PhysRevLett.69.2863} {\bibfield
  {journal} {\bibinfo  {journal} {Phys. Rev. Lett.}\ }\textbf {\bibinfo
  {volume} {69}},\ \bibinfo {pages} {2863} (\bibinfo {year}
  {1992})}\BibitemShut {NoStop}%
\bibitem [{\citenamefont {Schollw{\"o}ck}(2011)}]{Schollwoeck11}%
  \BibitemOpen
  \bibfield  {author} {\bibinfo {author} {\bibfnamefont {U.}~\bibnamefont
  {Schollw{\"o}ck}},\ }\href {\doibase 10.1016/j.aop.2010.09.012} {\bibfield
  {journal} {\bibinfo  {journal} {Ann. Phys.}\ }\textbf {\bibinfo {volume}
  {326}},\ \bibinfo {pages} {96} (\bibinfo {year} {2011})}\BibitemShut
  {NoStop}%
\bibitem [{\citenamefont {Weichselbaum}(2012{\natexlab{a}})}]{Wb12}%
  \BibitemOpen
  \bibfield  {author} {\bibinfo {author} {\bibfnamefont {A.}~\bibnamefont
  {Weichselbaum}},\ }\href {\doibase https://doi.org/10.1016/j.aop.2012.07.009}
  {\bibfield  {journal} {\bibinfo  {journal} {Annals of Physics}\ }\textbf
  {\bibinfo {volume} {327}},\ \bibinfo {pages} {2972 } (\bibinfo {year}
  {2012}{\natexlab{a}})}\BibitemShut {NoStop}%
\bibitem [{\citenamefont {Weichselbaum}(2012{\natexlab{b}})}]{Wb12tns}%
  \BibitemOpen
  \bibfield  {author} {\bibinfo {author} {\bibfnamefont {A.}~\bibnamefont
  {Weichselbaum}},\ }\href {\doibase 10.1103/PhysRevB.86.245124} {\bibfield
  {journal} {\bibinfo  {journal} {Phys. Rev. B}\ }\textbf {\bibinfo {volume}
  {86}},\ \bibinfo {pages} {245124} (\bibinfo {year}
  {2012}{\natexlab{b}})}\BibitemShut {NoStop}%
\bibitem [{\citenamefont {G\"uttge}\ \emph {et~al.}(2013)\citenamefont
  {G\"uttge}, \citenamefont {Anders}, \citenamefont {Schollw\"ock},
  \citenamefont {Eidelstein},\ and\ \citenamefont {Schiller}}]{Guettge13}%
  \BibitemOpen
  \bibfield  {author} {\bibinfo {author} {\bibfnamefont {F.}~\bibnamefont
  {G\"uttge}}, \bibinfo {author} {\bibfnamefont {F.~B.}\ \bibnamefont
  {Anders}}, \bibinfo {author} {\bibfnamefont {U.}~\bibnamefont
  {Schollw\"ock}}, \bibinfo {author} {\bibfnamefont {E.}~\bibnamefont
  {Eidelstein}}, \ and\ \bibinfo {author} {\bibfnamefont {A.}~\bibnamefont
  {Schiller}},\ }\href {\doibase 10.1103/PhysRevB.87.115115} {\bibfield
  {journal} {\bibinfo  {journal} {Phys. Rev. B}\ }\textbf {\bibinfo {volume}
  {87}},\ \bibinfo {pages} {115115} (\bibinfo {year} {2013})}\BibitemShut
  {NoStop}%
\bibitem [{\citenamefont {Verstraete}\ \emph {et~al.}(2004)\citenamefont
  {Verstraete}, \citenamefont {Garc\'{\i}a-Ripoll},\ and\ \citenamefont
  {Cirac}}]{Verstraete04}%
  \BibitemOpen
  \bibfield  {author} {\bibinfo {author} {\bibfnamefont {F.}~\bibnamefont
  {Verstraete}}, \bibinfo {author} {\bibfnamefont {J.~J.}\ \bibnamefont
  {Garc\'{\i}a-Ripoll}}, \ and\ \bibinfo {author} {\bibfnamefont {J.~I.}\
  \bibnamefont {Cirac}},\ }\href {\doibase 10.1103/PhysRevLett.93.207204}
  {\bibfield  {journal} {\bibinfo  {journal} {Phys. Rev. Lett.}\ }\textbf
  {\bibinfo {volume} {93}},\ \bibinfo {pages} {207204} (\bibinfo {year}
  {2004})}\BibitemShut {NoStop}%
\bibitem [{\citenamefont {Albert}\ and\ \citenamefont
  {Jiang}(2014)}]{Albert14}%
  \BibitemOpen
  \bibfield  {author} {\bibinfo {author} {\bibfnamefont {V.~V.}\ \bibnamefont
  {Albert}}\ and\ \bibinfo {author} {\bibfnamefont {L.}~\bibnamefont {Jiang}},\
  }\href {\doibase 10.1103/PhysRevA.89.022118} {\bibfield  {journal} {\bibinfo
  {journal} {Phys. Rev. A}\ }\textbf {\bibinfo {volume} {89}},\ \bibinfo
  {pages} {022118} (\bibinfo {year} {2014})}\BibitemShut {NoStop}%
\bibitem [{\citenamefont {Werner}\ \emph {et~al.}(2016)\citenamefont {Werner},
  \citenamefont {Jaschke}, \citenamefont {Silvi}, \citenamefont {Kliesch},
  \citenamefont {Calarco}, \citenamefont {Eisert},\ and\ \citenamefont
  {Montangero}}]{Werner16}%
  \BibitemOpen
  \bibfield  {author} {\bibinfo {author} {\bibfnamefont {A.~H.}\ \bibnamefont
  {Werner}}, \bibinfo {author} {\bibfnamefont {D.}~\bibnamefont {Jaschke}},
  \bibinfo {author} {\bibfnamefont {P.}~\bibnamefont {Silvi}}, \bibinfo
  {author} {\bibfnamefont {M.}~\bibnamefont {Kliesch}}, \bibinfo {author}
  {\bibfnamefont {T.}~\bibnamefont {Calarco}}, \bibinfo {author} {\bibfnamefont
  {J.}~\bibnamefont {Eisert}}, \ and\ \bibinfo {author} {\bibfnamefont
  {S.}~\bibnamefont {Montangero}},\ }\href {\doibase
  10.1103/PhysRevLett.116.237201} {\bibfield  {journal} {\bibinfo  {journal}
  {Phys. Rev. Lett.}\ }\textbf {\bibinfo {volume} {116}},\ \bibinfo {pages}
  {237201} (\bibinfo {year} {2016})}\BibitemShut {NoStop}%
\bibitem [{\citenamefont {Kraus}(1971)}]{KRAUS1971311}%
  \BibitemOpen
  \bibfield  {author} {\bibinfo {author} {\bibfnamefont {K.}~\bibnamefont
  {Kraus}},\ }\href {\doibase https://doi.org/10.1016/0003-4916(71)90108-4}
  {\bibfield  {journal} {\bibinfo  {journal} {Annals of Physics}\ }\textbf
  {\bibinfo {volume} {64}},\ \bibinfo {pages} {311 } (\bibinfo {year}
  {1971})}\BibitemShut {NoStop}%
\bibitem [{\citenamefont {Nielsen}\ and\ \citenamefont
  {Chuang}(2010)}]{nielsen_chuang_2010}%
  \BibitemOpen
  \bibfield  {author} {\bibinfo {author} {\bibfnamefont {M.~A.}\ \bibnamefont
  {Nielsen}}\ and\ \bibinfo {author} {\bibfnamefont {I.~L.}\ \bibnamefont
  {Chuang}},\ }\href {\doibase 10.1017/CBO9780511976667} {\emph {\bibinfo
  {title} {Quantum Computation and Quantum Information: 10th Anniversary
  Edition}}}\ (\bibinfo  {publisher} {Cambridge University Press},\ \bibinfo
  {year} {2010})\BibitemShut {NoStop}%
\bibitem [{\citenamefont {Corboz}\ \emph {et~al.}(2010)\citenamefont {Corboz},
  \citenamefont {Or{\'u}s}, \citenamefont {Bauer},\ and\ \citenamefont
  {Vidal}}]{Corboz10}%
  \BibitemOpen
  \bibfield  {author} {\bibinfo {author} {\bibfnamefont {P.}~\bibnamefont
  {Corboz}}, \bibinfo {author} {\bibfnamefont {R.}~\bibnamefont {Or{\'u}s}},
  \bibinfo {author} {\bibfnamefont {B.}~\bibnamefont {Bauer}}, \ and\ \bibinfo
  {author} {\bibfnamefont {G.}~\bibnamefont {Vidal}},\ }\href {\doibase
  10.1103/PhysRevB.81.165104} {\bibfield  {journal} {\bibinfo  {journal} {Phys.
  Rev. B}\ }\textbf {\bibinfo {volume} {81}},\ \bibinfo {pages} {165104}
  (\bibinfo {year} {2010})}\BibitemShut {NoStop}%
\bibitem [{\citenamefont {Goldstein}\ and\ \citenamefont
  {Berkovits}(2007)}]{Goldstein_2007}%
  \BibitemOpen
  \bibfield  {author} {\bibinfo {author} {\bibfnamefont {M.}~\bibnamefont
  {Goldstein}}\ and\ \bibinfo {author} {\bibfnamefont {R.}~\bibnamefont
  {Berkovits}},\ }\href {\doibase 10.1088/1367-2630/9/5/118} {\bibfield
  {journal} {\bibinfo  {journal} {New Journal of Physics}\ }\textbf {\bibinfo
  {volume} {9}},\ \bibinfo {pages} {118} (\bibinfo {year} {2007})}\BibitemShut
  {NoStop}%
\bibitem [{\citenamefont {Karrasch}\ \emph
  {et~al.}(2007{\natexlab{a}})\citenamefont {Karrasch}, \citenamefont {Hecht},
  \citenamefont {Weichselbaum}, \citenamefont {von Delft}, \citenamefont
  {Oreg},\ and\ \citenamefont {Meden}}]{Karrasch_2007}%
  \BibitemOpen
  \bibfield  {author} {\bibinfo {author} {\bibfnamefont {C.}~\bibnamefont
  {Karrasch}}, \bibinfo {author} {\bibfnamefont {T.}~\bibnamefont {Hecht}},
  \bibinfo {author} {\bibfnamefont {A.}~\bibnamefont {Weichselbaum}}, \bibinfo
  {author} {\bibfnamefont {J.}~\bibnamefont {von Delft}}, \bibinfo {author}
  {\bibfnamefont {Y.}~\bibnamefont {Oreg}}, \ and\ \bibinfo {author}
  {\bibfnamefont {V.}~\bibnamefont {Meden}},\ }\href {\doibase
  10.1088/1367-2630/9/5/123} {\bibfield  {journal} {\bibinfo  {journal} {New
  Journal of Physics}\ }\textbf {\bibinfo {volume} {9}},\ \bibinfo {pages}
  {123} (\bibinfo {year} {2007}{\natexlab{a}})}\BibitemShut {NoStop}%
\bibitem [{\citenamefont {Meir}\ and\ \citenamefont {Wingreen}(1992)}]{Meir92}%
  \BibitemOpen
  \bibfield  {author} {\bibinfo {author} {\bibfnamefont {Y.}~\bibnamefont
  {Meir}}\ and\ \bibinfo {author} {\bibfnamefont {N.~S.}\ \bibnamefont
  {Wingreen}},\ }\href {\doibase 10.1103/PhysRevLett.68.2512} {\bibfield
  {journal} {\bibinfo  {journal} {Phys. Rev. Lett.}\ }\textbf {\bibinfo
  {volume} {68}},\ \bibinfo {pages} {2512} (\bibinfo {year}
  {1992})}\BibitemShut {NoStop}%
\bibitem [{\citenamefont {Weichselbaum}\ and\ \citenamefont {von
  Delft}(2007)}]{Wb07}%
  \BibitemOpen
  \bibfield  {author} {\bibinfo {author} {\bibfnamefont {A.}~\bibnamefont
  {Weichselbaum}}\ and\ \bibinfo {author} {\bibfnamefont {J.}~\bibnamefont {von
  Delft}},\ }\href {\doibase 10.1103/PhysRevLett.99.076402} {\bibfield
  {journal} {\bibinfo  {journal} {Phys. Rev. Lett.}\ }\textbf {\bibinfo
  {volume} {99}},\ \bibinfo {pages} {076402} (\bibinfo {year}
  {2007})}\BibitemShut {NoStop}%
\bibitem [{\citenamefont {las Cuevas}\ \emph {et~al.}(2013)\citenamefont {las
  Cuevas}, \citenamefont {Schuch}, \citenamefont {P{\'{e}}rez-Garc{\'{\i}}a},\
  and\ \citenamefont {Cirac}}]{Cuevas_2013}%
  \BibitemOpen
  \bibfield  {author} {\bibinfo {author} {\bibfnamefont {G.~D.}\ \bibnamefont
  {las Cuevas}}, \bibinfo {author} {\bibfnamefont {N.}~\bibnamefont {Schuch}},
  \bibinfo {author} {\bibfnamefont {D.}~\bibnamefont
  {P{\'{e}}rez-Garc{\'{\i}}a}}, \ and\ \bibinfo {author} {\bibfnamefont
  {J.~I.}\ \bibnamefont {Cirac}},\ }\href {\doibase
  10.1088/1367-2630/15/12/123021} {\ \textbf {\bibinfo {volume} {15}},\
  \bibinfo {pages} {123021} (\bibinfo {year} {2013})}\BibitemShut {NoStop}%
\bibitem [{\citenamefont {Hauschild}\ \emph {et~al.}(2018)\citenamefont
  {Hauschild}, \citenamefont {Leviatan}, \citenamefont {Bardarson},
  \citenamefont {Altman}, \citenamefont {Zaletel},\ and\ \citenamefont
  {Pollmann}}]{Hauschild18}%
  \BibitemOpen
  \bibfield  {author} {\bibinfo {author} {\bibfnamefont {J.}~\bibnamefont
  {Hauschild}}, \bibinfo {author} {\bibfnamefont {E.}~\bibnamefont {Leviatan}},
  \bibinfo {author} {\bibfnamefont {J.~H.}\ \bibnamefont {Bardarson}}, \bibinfo
  {author} {\bibfnamefont {E.}~\bibnamefont {Altman}}, \bibinfo {author}
  {\bibfnamefont {M.~P.}\ \bibnamefont {Zaletel}}, \ and\ \bibinfo {author}
  {\bibfnamefont {F.}~\bibnamefont {Pollmann}},\ }\href {\doibase
  10.1103/PhysRevB.98.235163} {\bibfield  {journal} {\bibinfo  {journal} {Phys.
  Rev. B}\ }\textbf {\bibinfo {volume} {98}},\ \bibinfo {pages} {235163}
  (\bibinfo {year} {2018})}\BibitemShut {NoStop}%
\bibitem [{\citenamefont {Krumnow}\ \emph {et~al.}(2019)\citenamefont
  {Krumnow}, \citenamefont {Eisert},\ and\ \citenamefont {Legeza}}]{krumnow19}%
  \BibitemOpen
  \bibfield  {author} {\bibinfo {author} {\bibfnamefont {C.}~\bibnamefont
  {Krumnow}}, \bibinfo {author} {\bibfnamefont {J.}~\bibnamefont {Eisert}}, \
  and\ \bibinfo {author} {\bibfnamefont {O.}~\bibnamefont {Legeza}},\
  }\href@noop {} {\  (\bibinfo {year} {2019})},\ \Eprint
  {http://arxiv.org/abs/1904.11999} {arXiv:1904.11999 [cond-mat.stat-mech]}
  \BibitemShut {NoStop}%
\bibitem [{\citenamefont {Rams}\ and\ \citenamefont {Zwolak}(2020)}]{Rams20}%
  \BibitemOpen
  \bibfield  {author} {\bibinfo {author} {\bibfnamefont {M.~M.}\ \bibnamefont
  {Rams}}\ and\ \bibinfo {author} {\bibfnamefont {M.}~\bibnamefont {Zwolak}},\
  }\href {\doibase 10.1103/PhysRevLett.124.137701} {\bibfield  {journal}
  {\bibinfo  {journal} {Phys. Rev. Lett.}\ }\textbf {\bibinfo {volume} {124}},\
  \bibinfo {pages} {137701} (\bibinfo {year} {2020})}\BibitemShut {NoStop}%
\bibitem [{\citenamefont {Cui}\ \emph {et~al.}(2015)\citenamefont {Cui},
  \citenamefont {Cirac},\ and\ \citenamefont {Ba\~nuls}}]{Cui15}%
  \BibitemOpen
  \bibfield  {author} {\bibinfo {author} {\bibfnamefont {J.}~\bibnamefont
  {Cui}}, \bibinfo {author} {\bibfnamefont {J.~I.}\ \bibnamefont {Cirac}}, \
  and\ \bibinfo {author} {\bibfnamefont {M.~C.}\ \bibnamefont {Ba\~nuls}},\
  }\href {\doibase 10.1103/PhysRevLett.114.220601} {\bibfield  {journal}
  {\bibinfo  {journal} {Phys. Rev. Lett.}\ }\textbf {\bibinfo {volume} {114}},\
  \bibinfo {pages} {220601} (\bibinfo {year} {2015})}\BibitemShut {NoStop}%
\bibitem [{\citenamefont {Mascarenhas}\ \emph {et~al.}(2015)\citenamefont
  {Mascarenhas}, \citenamefont {Flayac},\ and\ \citenamefont
  {Savona}}]{Mascarenhas15}%
  \BibitemOpen
  \bibfield  {author} {\bibinfo {author} {\bibfnamefont {E.}~\bibnamefont
  {Mascarenhas}}, \bibinfo {author} {\bibfnamefont {H.}~\bibnamefont {Flayac}},
  \ and\ \bibinfo {author} {\bibfnamefont {V.}~\bibnamefont {Savona}},\ }\href
  {\doibase 10.1103/PhysRevA.92.022116} {\bibfield  {journal} {\bibinfo
  {journal} {Phys. Rev. A}\ }\textbf {\bibinfo {volume} {92}},\ \bibinfo
  {pages} {022116} (\bibinfo {year} {2015})}\BibitemShut {NoStop}%
\bibitem [{\citenamefont {Gogolin}\ \emph {et~al.}(2004)\citenamefont
  {Gogolin}, \citenamefont {Nersesyan},\ and\ \citenamefont
  {Tsvelik}}]{gogolin2004bosonization}%
  \BibitemOpen
  \bibfield  {author} {\bibinfo {author} {\bibfnamefont {A.~O.}\ \bibnamefont
  {Gogolin}}, \bibinfo {author} {\bibfnamefont {A.~A.}\ \bibnamefont
  {Nersesyan}}, \ and\ \bibinfo {author} {\bibfnamefont {A.~M.}\ \bibnamefont
  {Tsvelik}},\ }\href@noop {} {\emph {\bibinfo {title} {Bosonization and
  strongly correlated systems}}}\ (\bibinfo  {publisher} {Cambridge university
  press},\ \bibinfo {year} {2004})\BibitemShut {NoStop}%
\bibitem [{\citenamefont {Karrasch}\ \emph
  {et~al.}(2007{\natexlab{b}})\citenamefont {Karrasch}, \citenamefont {Hecht},
  \citenamefont {Weichselbaum}, \citenamefont {Oreg}, \citenamefont {von
  Delft},\ and\ \citenamefont {Meden}}]{Karrasch07PRL}%
  \BibitemOpen
  \bibfield  {author} {\bibinfo {author} {\bibfnamefont {C.}~\bibnamefont
  {Karrasch}}, \bibinfo {author} {\bibfnamefont {T.}~\bibnamefont {Hecht}},
  \bibinfo {author} {\bibfnamefont {A.}~\bibnamefont {Weichselbaum}}, \bibinfo
  {author} {\bibfnamefont {Y.}~\bibnamefont {Oreg}}, \bibinfo {author}
  {\bibfnamefont {J.}~\bibnamefont {von Delft}}, \ and\ \bibinfo {author}
  {\bibfnamefont {V.}~\bibnamefont {Meden}},\ }\href {\doibase
  10.1103/PhysRevLett.98.186802} {\bibfield  {journal} {\bibinfo  {journal}
  {Phys. Rev. Lett.}\ }\textbf {\bibinfo {volume} {98}},\ \bibinfo {pages}
  {186802} (\bibinfo {year} {2007}{\natexlab{b}})}\BibitemShut {NoStop}%
\bibitem [{\citenamefont {Goldstein}\ \emph {et~al.}(2010)\citenamefont
  {Goldstein}, \citenamefont {Berkovits},\ and\ \citenamefont
  {Gefen}}]{Goldstein10}%
  \BibitemOpen
  \bibfield  {author} {\bibinfo {author} {\bibfnamefont {M.}~\bibnamefont
  {Goldstein}}, \bibinfo {author} {\bibfnamefont {R.}~\bibnamefont
  {Berkovits}}, \ and\ \bibinfo {author} {\bibfnamefont {Y.}~\bibnamefont
  {Gefen}},\ }\href {\doibase 10.1103/PhysRevLett.104.226805} {\bibfield
  {journal} {\bibinfo  {journal} {Phys. Rev. Lett.}\ }\textbf {\bibinfo
  {volume} {104}},\ \bibinfo {pages} {226805} (\bibinfo {year}
  {2010})}\BibitemShut {NoStop}%
\bibitem [{\citenamefont {Sakai}\ and\ \citenamefont
  {Kuramoto}(1994)}]{Sakai94}%
  \BibitemOpen
  \bibfield  {author} {\bibinfo {author} {\bibfnamefont {O.}~\bibnamefont
  {Sakai}}\ and\ \bibinfo {author} {\bibfnamefont {Y.}~\bibnamefont
  {Kuramoto}},\ }\href {\doibase https://doi.org/10.1016/0038-1098(94)90589-4}
  {\bibfield  {journal} {\bibinfo  {journal} {Solid State Communications}\
  }\textbf {\bibinfo {volume} {89}},\ \bibinfo {pages} {307 } (\bibinfo {year}
  {1994})}\BibitemShut {NoStop}%
\bibitem [{\citenamefont {Bulla}\ \emph {et~al.}(1998)\citenamefont {Bulla},
  \citenamefont {Hewson},\ and\ \citenamefont {Pruschke}}]{Bulla98}%
  \BibitemOpen
  \bibfield  {author} {\bibinfo {author} {\bibfnamefont {R.}~\bibnamefont
  {Bulla}}, \bibinfo {author} {\bibfnamefont {A.~C.}\ \bibnamefont {Hewson}}, \
  and\ \bibinfo {author} {\bibfnamefont {T.}~\bibnamefont {Pruschke}},\ }\href
  {\doibase 10.1088/0953-8984/10/37/021} {\bibfield  {journal} {\bibinfo
  {journal} {Journal of Physics: Condensed Matter}\ }\textbf {\bibinfo {volume}
  {10}},\ \bibinfo {pages} {8365} (\bibinfo {year} {1998})}\BibitemShut
  {NoStop}%
\bibitem [{\citenamefont {Stadler}\ \emph {et~al.}(2015)\citenamefont
  {Stadler}, \citenamefont {Yin}, \citenamefont {von Delft}, \citenamefont
  {Kotliar},\ and\ \citenamefont {Weichselbaum}}]{Stadler15}%
  \BibitemOpen
  \bibfield  {author} {\bibinfo {author} {\bibfnamefont {K.~M.}\ \bibnamefont
  {Stadler}}, \bibinfo {author} {\bibfnamefont {Z.~P.}\ \bibnamefont {Yin}},
  \bibinfo {author} {\bibfnamefont {J.}~\bibnamefont {von Delft}}, \bibinfo
  {author} {\bibfnamefont {G.}~\bibnamefont {Kotliar}}, \ and\ \bibinfo
  {author} {\bibfnamefont {A.}~\bibnamefont {Weichselbaum}},\ }\href {\doibase
  10.1103/PhysRevLett.115.136401} {\bibfield  {journal} {\bibinfo  {journal}
  {Phys. Rev. Lett.}\ }\textbf {\bibinfo {volume} {115}},\ \bibinfo {pages}
  {136401} (\bibinfo {year} {2015})}\BibitemShut {NoStop}%
\bibitem [{\citenamefont {Oliveira}\ and\ \citenamefont
  {Oliveira}(1994)}]{Oliveira94}%
  \BibitemOpen
  \bibfield  {author} {\bibinfo {author} {\bibfnamefont {W.~C.}\ \bibnamefont
  {Oliveira}}\ and\ \bibinfo {author} {\bibfnamefont {L.~N.}\ \bibnamefont
  {Oliveira}},\ }\href {\doibase 10.1103/PhysRevB.49.11986} {\bibfield
  {journal} {\bibinfo  {journal} {Phys. Rev. B}\ }\textbf {\bibinfo {volume}
  {49}},\ \bibinfo {pages} {11986} (\bibinfo {year} {1994})}\BibitemShut
  {NoStop}%
\bibitem [{\citenamefont {Calabrese}\ and\ \citenamefont
  {Lefevre}(2008)}]{Calabrese08}%
  \BibitemOpen
  \bibfield  {author} {\bibinfo {author} {\bibfnamefont {P.}~\bibnamefont
  {Calabrese}}\ and\ \bibinfo {author} {\bibfnamefont {A.}~\bibnamefont
  {Lefevre}},\ }\href {\doibase 10.1103/PhysRevA.78.032329} {\bibfield
  {journal} {\bibinfo  {journal} {Phys. Rev. A}\ }\textbf {\bibinfo {volume}
  {78}},\ \bibinfo {pages} {032329} (\bibinfo {year} {2008})}\BibitemShut
  {NoStop}%
\end{thebibliography}%

\end{document}